\documentclass[twocolumn,pra,amsmath,amssymb]{revtex4-1}
\usepackage{graphicx,bm,epsfig,color}

\begin{document}

\title{Consequences of unitary evolution of coupled qubit-resonator systems\\
for stabilizing circuits in surface codes}

\author{H.W.L. Naus} \email{rik.naus@tno.nl}
\author{R. Versluis}\email{richard.versluis@tno.nl}
\affiliation{Quantum Technology, TNO, P.O. Box 155, 2600 AD Delft, The Netherlands and
QuTech, Delft University of Technology, P.O. Box 5046, 2600 GA Delft, The Netherlands}

\begin{abstract}
Surface codes based on stabilizer circuits may pave the way for large scale fault-tolerant
quantum computation.
The surface code, being a fully planar implementation of a larger family of error correction codes,
uses only single- and two-qubit gates and the error threshold falls close to $1\%$ for a large range of errors.
Among the most promising candidates to physically implement such circuits and codes are
superconducting qubits,
such as transmons, coupled by resonators to enable the two-qubit interactions.
In this study, we investigate
a $X$ and $Z$ stabilizing circuit realized by two data qubits, two ancillas and four resonators.
The aim is to assess the consequences of unitary evolution of the interacting system, in particular
for given stable initial states, on the fidelity of the output states and the probabilities of obtaining
the correct error syndrome (capture probabilities).
To this end, we model the system with a Jaynes-Tavis-Cummings Hamiltonian and construct
the low-excitation level evolution operators.
The analysis is limited to two (out of four) stable input states.
We assume an ideal system with perfect single- and two-qubit gates, perfect measurements taking zero measurement time
and no decoherence or leakage. 
Our analysis shows that the capture probabilities after the execution of a single stabilizer
round are not equal to $100\%$, but vary between $99.2\%$ and $99.99\%$. 
This is caused solely by the unitary evolution of the interacting system. 
Two consecutive rounds of stabilizer measurements result in capture probability values that depend
heavily on the duration of the evolution, but vary between $0\%$ and $99\%$. 
Also due to the unitary evolution, the final state of the data qubits leaves the the four-dimensional subspace,
which results in a state fidelity oscillating between $0$ and $1$.
Even if an error on the qubits is captured, the correcting operation on the qubit will not bring the
qubit to the original state.
The errors induced by the Hamiltonian evolution
of the system cannot be interpreted nor classified as commonly appearing errors.
Additional or augmented quantum error correction is possibly required to compensate these effects of
resonator-qubit interaction. 
\end{abstract}

\date{\today}
\maketitle

\newcommand{\ketz}{|\,\mathbf{0}\rangle}
\newcommand{\keto}{|\,\mathbf{1}\rangle}
\newcommand{\braz}{\langle\mathbf{0}\,|}
\newcommand{\brao}{\langle\mathbf{1}\,|}

\newcommand{\ketnz}{|\,n, \mathbf{0}\rangle}
\newcommand{\ketno}{|\,n, \mathbf{1}\rangle}
\newcommand{\branz}{\langle n, \mathbf{0}\,|}

\section{Introduction}
In order to perform computations with a gate-based universal quantum computer,
quantum error correction is mandatory \cite{Devitt}. Surface codes, operated as 
stabilizer codes can lead to large scale, fault-tolerant quantum computing \cite{Fowler,Brav,Terhal}.
The goal is to identify and correct several quantum errors meaning that qubits are
not in the state aimed for. Such errors can be caused by environmental decoherence, 
imperfect knowledge of the quantum system yielding errors in coherent control, imperfect
initialization and loss of qubits. Leakage to unwanted states
and measurement errors may happen as well. In principle, stabilizer codes indeed can correct
such uncorrelated errors. For quantum processors based on superconducting qubits, scalable
circuit and control have been recently proposed to execute the error-correction cycles \cite{Versluis}.

Superconducting quantum systems are promising candidates for near-term quantum computers \cite{Chen,Brien}. 
Such devices consist out of transmon(-like) qubits which are coupled
by resonators \cite{Carlo,Mavro}. This coupling is exploited to perform two-qubit gates  necessary 
for universal computing and error-correction codes. It is important to realize that
the qubit-resonator interaction cannot be switched off in the state-of-the-art quantum circuits.
It can be made effectively low by detuning the frequencies of qubit and resonators. 
The near-resonant situation is however necessary to obtain sufficient entanglement
for gate operations. 
Note that the coupling of read-out resonators or cavities can be tuned to low values
necessary to obtain the dispersive limit, enabling quantum non-demolition measurements.
The situation for electron spin qubits in quantum dots is somewhat different since the exchange
interaction between the electrons can be tuned to almost arbitrarily small values. 

In this paper, we study the unitary evolution of a four-qubit system coupled by four resonators
in the context of a stabilizing circuit. Our specific choice is motivated by the four-qubit
$X$ and $Z$ stabilizing circuit proposed in \cite{Fowler}; see also \cite{AC}.
In contrast with \cite{Fowler}, we take into account the unitary
 evolution of the interacting system of qubits and resonators.
The stable two-qubit states are the well-known Bell states.
We consider as initial state of the complete system --
two data qubits, two ancillas and resonators --  a product state of such a Bell state, the corresponding two-ancilla
state and empty resonators. The unitary evolution of this state governed by 
a Jaynes-Tavis-Cummings Hamiltonian is calculated with the techniques of \cite{Qudit}.
Consequences for subsequent measurement probabilities and state fidelities
are assessed.

Our work is related to the recent gate-error analysis \cite{gate-error}, where the dynamics of
two transmons coupled by a resonator has been studied in relation to gates implemented by pulses.
Systematic errors due to the usually neglected interaction of the total system have been identified.
Note that the unitary evolution yields a `detrimental effect' on the computational
subspace. We will obtain similar results in the context of a stabilizer circuit and show
that the fidelity of a desired (Bell) state in the two data qubit space detoriates rapidly.
These errors cannot be corrected by the stabilizer code under consideration. Recently, a study 
addressing fault-tolerance by including dynamics in small systems has appeared \cite{Will}.

The outline of this paper is as follows.
The next section presents the $X$ and $Z$ stabilizing circuit as well as the corresponding physical
implementation in terms of data qubits, ancillas and resonators. It also explicity defines the model
Hamiltonian and solutions for the lowest excitation subspaces. Subsequently, the unitary evolution of
two complete initial states, each containing a stable Bell state, is studied in \ref{sec:UEBM}.
Corresponding fidelities
and re-insertion of the final state into the stabilizing circuit are investigated in the sections
\ref{sec:FIDEC} and \ref{sec:RINS}.
Unitary free evolution and the related transformation to a rotating frame are addressed in \ref{sec:FRF}.
It is followed in section \ref{sec:THEO} by a theoretical discussion assessing results  and consequences in the context of error
analysis, stabilizers and quantum operations. The last section \ref{sec:CONC} contains conclusions.
Note that some adopted definitions and useful relations are given in appendix \ref{app:A}.

\section{$X$ and $Z$ stabilizing circuit}\label{sec:XZcircuit}
\begin{figure}[htb]
\centering
\includegraphics[width=210pt]{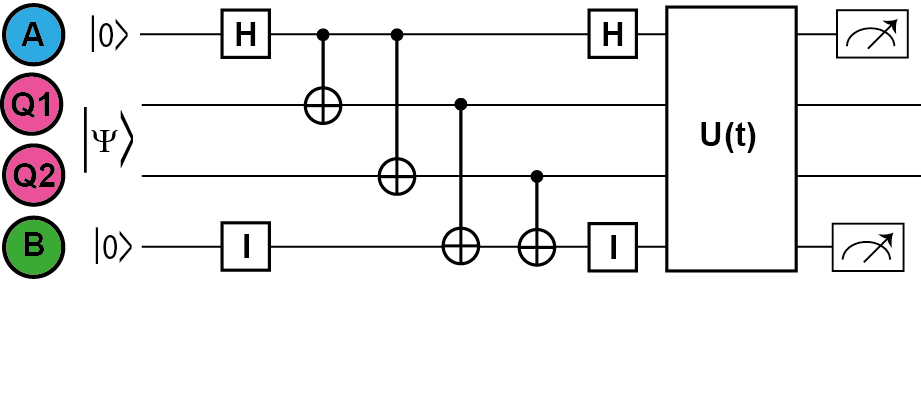}
\caption{\label{fig:circuit} Quantum $X$ and $Z$ stabilizer circuit including unitary evolution; the ideal circuit
does not include $U(t)$.}
\end{figure}

We consider the $X$ and $Z$ stabilizer circuit as presented in appendix B
of \cite{Fowler}, however extended with unitary (Hamiltonian) evolution $U(t)$
for time duration $t$, cf.  Figure \ref{fig:circuit}.
The ideal quantum circuit does not include $U$ and is part of the
in \cite{Fowler,Brav,Terhal} outlined surface code.
It is  built with
two data qubits and two ancillas, also called measure qubits. The corresponding 
measurement result is the {\em error syndrome} \cite{NC}.

\subsection{Circuit analysis}
The ideal circuit is analyzed by \cite{Fowler} in terms of the standard (computational)  basis.
For convenience, we repeat the analysis in terms of the Bell states in appendix \ref{app:appB}.
In both ways it is demonstrated that
the two data qubits are stabilized in a simultaneous eigenstate of
$X^{[1]} X^{[2]}$ and $Z^{[1]} Z^{[2]}$. These are exactly the Bell states:
\begin{eqnarray}
X^{[1]} X^{[2]} |\Phi^+ \rangle =  |\Phi^+ \rangle, &\quad&
X^{[1]} X^{[2]} |\Psi^+ \rangle =  |\Psi^+ \rangle, \nonumber \\
X^{[1]} X^{[2]} |\Phi^- \rangle =  -|\Phi^- \rangle, &\quad&
X^{[1]} X^{[2]} |\Psi^- \rangle =  -|\Psi^- \rangle, \nonumber \\
Z^{[1]} Z^{[2]} |\Phi^+ \rangle =  |\Phi^+ \rangle, &\quad&
Z^{[1]} Z^{[2]} |\Psi^+ \rangle = - |\Psi^+ \rangle, \nonumber \\
Z^{[1]} Z^{[2]} |\Phi^- \rangle =  |\Phi^- \rangle, &\quad&
Z^{[1]} Z^{[2]} |\Psi^- \rangle = - |\Psi^- \rangle.
\end{eqnarray}
In other words, if we select one of these states as input to the circuit,
it can readily be verified that
the same output state state is obtained. The result of the ancilla measurement, the error syndrome,
is reproduced as well with probability one.
Thus we have shown that this indeed is a stabilizer circuit.

\subsection{Error correction}\label{sec:EC}
Assume that we put a stable state into the circuit. As mentioned above,
without errors, the ancilla measurement result is known. Suppose, however, that
due to some error a different error syndrome is obtained. Concomitantly,
the resulting two-qubit state has changed as well. Because the final state
 is nevertheless known,
error correction by applying one or two single qubit operations is possible.
We work out this procedure for two out of four cases, thereby explicitly constructing the
necessary operations. Equations (\ref{eq:Xact}, \ref{eq:Zact}) are repeatedly used.
\begin{itemize}
\item{State $|\Phi^+\rangle$}\\
Error syndrome $(+1,+1)$, that is no errors are detected; otherwise: 
	\begin{itemize}
        \item Error syndrome $(-1,+1)$, state $|\Phi^-\rangle$, correction 
$Z^{[1]}|\Phi^-\rangle= |\Phi^+\rangle$.
        \item Error syndrome $(+1,-1)$, state $|\Psi^+\rangle$, correction
$X^{[2]}|\Psi^+\rangle= |\Phi^+\rangle$.
        \item Error syndrome $(-1,-1)$, state $|\Psi^-\rangle$, correction 
$ X^{[2]} Z^{[1]}|\Psi^-\rangle= |\Phi^+\rangle$.
	\end{itemize}
\item{State $|\Psi^+\rangle$}\\
Error syndrome $(+1,-1)$, that is no errors are detected; otherwise: 
	\begin{itemize}
        \item Error syndrome $(+1,+1)$, state $|\Phi^+\rangle$, correction  
$X^{[1]}|\Phi^+\rangle= |\Psi^+\rangle$.
        \item Error syndrome $(-1,+1)$, state $|\Phi^-\rangle$, correction
$X^{[2]} Z^{[1]}|\Phi^-\rangle= |\Psi^+\rangle$.
        \item Error syndrome $(-1,-1)$, state $|\Psi^-\rangle$, correction 
$ Z^{[1]}|\Psi^-\rangle= |\Psi^+\rangle$.
	\end{itemize}
\end{itemize}
Note that these error corrections are not unique, for example $Z^{[2]}|\Phi^-\rangle=|\Phi^+\rangle$.

\subsection{Dynamics}
\begin{figure}[htb]
\centering
\includegraphics[width=110pt]{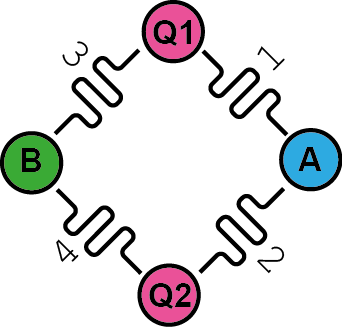}
\caption{\label{fig:system} Two data qubits and two ancillas coupled by four resonators.}
\end{figure}

Two data qubits stabilized by one measure-$X$ qubit $A$ and one measure-$Z$ qubit $B$ \cite{Fowler}
coupled by resonators are shown and labelled in Figure \ref{fig:system}.

The generalized Jaynes-Tavis-Cummings Hamiltonian for this system is given by \cite{Qudit}
\begin{eqnarray}
H&=& H_0^q + H_0^r+ H_{\text{int}}^{11}+ H_{\text{int}}^{1a}+ H_{\text{int}}^{2a} + H_{\text{int}}^{22}\nonumber \\
&+& H_{\text{int}}^{31}+ H_{\text{int}}^{3b}+ H_{\text{int}}^{4b} + H_{\text{int}}^{42},
\label{eq:HJC}
\end{eqnarray}
with the `free' qubit Hamiltonian
\begin{equation}
H_0^q= -\tfrac{1}{2} \omega_1' \sigma_z^{[1]}
-\tfrac{1}{2} \omega_a \sigma_z^{[a]} -\tfrac{1}{2} \omega_2' \sigma_z^{[2]}
-\tfrac{1}{2} \omega_b \sigma_z^{[b]} 
\end{equation}
and the Hamiltonian of the resonators
\begin{eqnarray}
H_0^r &=& 
 \omega_1(a_1^\dagger a_1 + \tfrac{1}{2})
+ \omega_2(a_2^\dagger a_2 + \tfrac{1}{2}) \nonumber \\
 &+& \omega_3(a_3^\dagger a_3 + \tfrac{1}{2})
+ \omega_4(a_4^\dagger a_4 + \tfrac{1}{2}).
\end{eqnarray}
The interaction terms in the rotating wave approximation read
\begin{eqnarray}
H_{\text{int}}^{11} &=& g_{11}\left(a_1^\dagger \sigma_+^{[1]}+a_1\sigma_-^{[1]}\right), \nonumber\\
H_{\text{int}}^{1a} &=& g_{1a}\left(a_1^\dagger \sigma_+^{[a]}+a_1\sigma_-^{[a]}\right), \quad \text{etc}.
\end{eqnarray}
The frequencies
are denoted by $\omega_k, k=1,\dots, 4,$  for the resonators, $\omega'_1, \omega'_2$
for the data qubits and $\omega_a, \omega_b$ for the ancillas. The coupling constants
are taken as $g_{11}, g_{1a}, g_{2a}, g_{22}, g_{42}, g_{4b}, g_{3b}, g_{31}$.

To study the dynamics, it is necessary to solve the Schr\"odinger equation, that is 
obtaining eigenenergies and eigenstates of the Hamiltonian. Herewith the
evolution operator $U(t)$ is constructed. The technique of \cite{Qudit} is applied, 
where excitation number operators for Jaynes-Tavis-Cummings Hamiltonians are exploited.
For the system under consideration this operator $\mathcal{N}$ is given by 
\begin{equation}
  \mathcal N = \sum_{k=1}^4 a_k^\dagger a_k - \tfrac{1}{2}(\sigma_z^{[1]}
  +\sigma_z^{[2]}+ \sigma_z^{[a]}+ \sigma_z^{[b]}).
\end{equation}
It commutes with the Hamiltonian:
$[\mathcal N, H] = 0$. Consequently, it is conserved  and $\mathcal N$ and $H$
can be diagonalized simultaneously.
Thus the dynamics can be analyzed separately for each excitation level.

The eigenstates of the excitation number operator are product states
$|r_1, r_2, r_3, r_4 \rangle \otimes |q_1, q_2 \rangle \otimes |q_a,q_b\rangle$,
with $r_k=0,1,2,\dots$ for $k=1,2,3,4$ and $q_j=\bm{0,1}$ for $j=1,2,a,b$.
The lowest level $\mathcal N$ state is also the ground state of the Hamiltonian  
\begin{equation}
|E_0\rangle = |0,0,0,0\rangle \otimes |\bm{0},\bm{0}\rangle \otimes |\bm{0}, \bm{0}\rangle,
\end{equation}
with energy eigenvalue
\begin{equation}
E_0 = \tfrac{1}{2}(\omega_1+\omega_2+\omega_3 +\omega_4-\omega'_1-\omega'_2-\omega_a-\omega_b).
\end{equation}
The first excitation level is eight-fold degenerate with respect to $\mathcal N$;
we choose the basis states as given in appendix \ref{app:appC}.
The matrix elements of the Hamiltonian in this subspace $\mathcal H_{ij}, i,j =1,2,\dots 8$
are also listed in appendix \ref{app:appC}.
The corresponding eigenvalue problem needs to be solved:
\begin{equation}
\mathcal H a_\mu = E_{1\mu}  a_\mu, \quad \mu=1,2,\dots,8.
\end{equation} As in \cite{Qudit}, this is done
by means of the Jacobi method.
In this way, we obtain the eigenstates
\begin{equation}
| E_{1\mu} \rangle = \sum_{k=1}^8 a_{\mu k} |e_k \rangle
\end{equation}
and the subspace evolution operator
\begin{equation}
U_1(t)= \sum_{\mu=1}^8 e^{-iE_{1\mu}t} |E_{1\mu} \rangle
\langle  E_{1\mu} | .
\end{equation}

We proceed to the second excitation level which has dimension thirty-two. See appendix \ref{app:appC}
for our choice  of basis states $|f_k\rangle$.
The matrix elements in the second excitation subspace 
$\mathsf{H}_{kl},  k,l=1,2, \dots, 32$
are  explicitly given in appendix \ref{app:appC} as well.
The eigenvalue problem in this subspace
\begin{equation}
\mathsf H \, b_\eta = E_{2\eta} b_\eta, \quad \eta=1,2,\dots,32,
\end{equation}
is again solved with the Jacobi method.
It yields the eigenstates
\begin{equation}
| E_{2\eta} \rangle = \sum_{l=1}^{32} b_{\eta l} |f_l \rangle
\end{equation}
and the second excitation subspace evolution operator
\begin{equation}
U_2(t)= \sum_{\eta=1}^{32} e^{-iE_{2\eta}t} |E_{2\eta} \rangle
\langle  E_{2\eta} |.
\end{equation}
The constructed evolution operators will be used to compute specific unitary evolutions.

\section{Unitary evolution}\label{sec:UEBM}
In this paper, we presuppose that the $X$ and $Z$ stabilizing circuit is ideal
up to and including the last Hadamard operation, step 7. in appendix \ref{app:appB}. 
If one of the stable Bell states (\ref{eq:Bell}) has been put in, 
it is reproduced in (\ref{eq:Had4}) because the corresponding coefficient
is one and the others are zero.  Before completing the circuit by measuring
the ancillas we now assume unitary evolution governed by the complete Hamiltonian
and analyze the consequences.
This is done for two initial states, containing the Bell states $|\Phi^+\rangle$ and $|\Psi^+\rangle$.
Assuming empty resonators, only the ground state and the second excitation subspace need to be
taken into account.
In principle, the analysis for the  states
 $|\Phi^-\rangle$ and $|\Psi^-\rangle$ proceeds analogously. The corresponding 
states, however, have components  in the third excitation subspace which makes
computations more cumbersome.  Therefore, we restrict ourselves to 
$|\Phi^+\rangle$ and $|\Psi^+\rangle$.

\subsection{Bell state $|\Phi^+\rangle$}
\begin{figure}[htb]
\centering
\includegraphics[width=210pt]{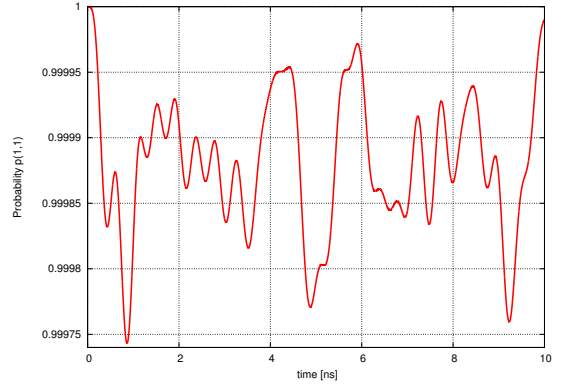}
\caption{\label{fig:41pp} Probability for obtaining the error syndrome (1,1)
after unitary evolution of the state $|0,0,0,0\rangle |\Phi^+\rangle |\mathbf{0}, \mathbf{0}\rangle$
as a function of time.
Parameters: 
$f_1=8.14, f_2=8.18, f_3=8.1, f_4=8.06, f_1'=6.6, f_2'=6.4, f_a=5.9, f_b=5.7$,
$g_{11}=0.51, g_{22}=0.52, g_{1a}=0.53, g_{2a}=0.54,  
g_{42}=0.5, g_{4b}=0.49, g_{3b}=0.48, g_{31}=0.47$; all values are given in GHz.}
\end{figure}

First, we assume that the two-qubit Bell state $|\Phi^+\rangle$
has been put in the ideal stabilizer circuit.
Then we study unitary evolution of the system after the last Hadamard operation
has been performed.  
Thus the `initial' state follows from (\ref{eq:Had4}) with $A_+=1, B_+=A_-=B_-=0$
and the inclusion with resonators, presumed to be empty
\begin{equation}
|\varphi_0\rangle = |0,0,0,0\rangle \otimes |\Phi^+\rangle \otimes 
|\mathbf{0}, \mathbf{0}\rangle.
\label{eq:ini}
\end{equation}
In terms of the defined basis vectors we get a superposition
of the ground state and a state in the second excitation subspace
\begin{equation}
|\varphi_0\rangle = \tfrac{1}{2}\sqrt{2}\left( |E_0\rangle + |f_{27}\rangle\right).
\end{equation}
Its unitary evolution is therefore given by
\begin{eqnarray}
|\varphi(t) \rangle &=& \left(U_0(t)+U_1(t)+U_2(t)+\dots \right)|\varphi_0\rangle \nonumber \\
&=& \tfrac{1}{2}\sqrt{2}\left( U(t) |E_0\rangle + U_2(t)|f_{27}\rangle \right) \nonumber \\
&=& \tfrac{1}{2}\sqrt{2}\left( e^{-iE_0t} |E_0\rangle + \sum_{\eta=1}^{32}
e^{-iE_{2\eta}t} b_{\eta 27} |E_{2\eta}\rangle \right) \nonumber \\
&=& \tfrac{1}{2}\sqrt{2}\left( e^{-iE_0t} |E_0\rangle + \sum_{\eta=1}^{32}
\sum_{k=1}^{32} e^{-iE_{2\eta}t} b_{\eta 27} b_{\eta k} | f_k\rangle \right) \nonumber \\
&=& \tfrac{1}{2}\sqrt{2} e^{-iE_0t} |E_0\rangle +
\sum_{k=1}^{32} \beta_{k}(t) | f_k\rangle,
\label{eq:euphi}
\end{eqnarray}
where we have defined
\begin{equation}
\beta_k(t) = \tfrac{1}{2}\sqrt{2} \sum_{\eta=1}^{32}
 e^{-iE_{2\eta}t} b_{\eta 27} b_{\eta k}.
\end{equation}
At this point we fix the time $t$ and perform the measurement on the ancillas.
The measurement operator corresponding to the error syndrome $(+1,+1)$ reads
\begin{equation}
P_{1,1} = \mathcal{I}_{r_1} \otimes \mathcal{I}_{r_2} \otimes
\mathcal{I}_{r_3} \otimes \mathcal{I}_{r_4}
 \otimes \mathcal{I}_{12} \otimes \ketz \braz \otimes \ketz \braz,
\end{equation}
with identity operators $\mathcal I_{r_j}, \mathcal I_{12}= \mathcal I^{[1]}\otimes \mathcal I^{[2]}$.
The probability that this result is obtained follows as
\begin{equation}
p(1,1)= \tfrac{1}{2} + \sum_{k \in S(1,1)} \beta_k^*(t) \beta_k(t),
\end{equation}
where the integer set is given as
\begin{equation}
S(1,1)= \{1,2,\cdots,12,15,16,19,20,23,24,27 \}.
\end{equation}
This result is depicted in Figure (\ref{fig:41pp}).

The state after projection and normalization follows as
\begin{equation}
|\varphi_{1,1}\rangle = \frac{1}{\sqrt{p(1,1)}} \left( \tfrac{1}{2}\sqrt{2} e^{-iE_0t} |E_0\rangle +
\sum_{k \in S(1,1)} \beta_{k}(t) | f_k\rangle \right).
\label{eq:PHIP}
\end{equation}
It can be written as a tensor product by factoring out the ancilla states
\begin{equation}
|\varphi_{1,1}\rangle = |\tilde \varphi_{1,1}\rangle \otimes
\ketz \otimes \ketz.
\label{eq:phipl11}
\end{equation}
The data qubits--resonators state, however, is still entangled.

\begin{figure}[htb]
\centering
\includegraphics[width=210pt]{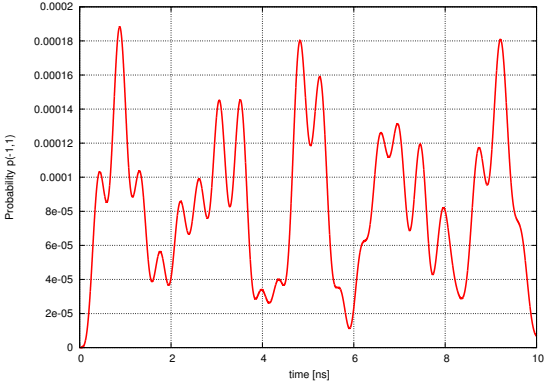}
\caption{\label{fig:41mp} Probability for obtaining the error syndrome (-1,1)
after unitary evolution of the state $|0,0,0,0\rangle |\Phi^+\rangle |\mathbf{0}, \mathbf{0}\rangle$
as a function of time.
Parameters as in the previous figure.}
\end{figure}

We continue with measurements which indicate an error in the state.
If the error syndrome equals $(-1,1)$, the corresponding measurement operator is given by
\begin{equation}
P_{-1,1} = \mathcal{I}_{r_1} \otimes \mathcal{I}_{r_2} \otimes
\mathcal{I}_{r_3} \otimes \mathcal{I}_{r_4}
 \otimes \mathcal{I}_{12} \otimes \keto \brao \otimes \ketz \braz.
\end{equation}
We obtain for the probability to find this result
\begin{equation}
p(-1,1)=  \sum_{k \in S(-1,1)} \beta_k^*(t) \beta_k(t),
\end{equation}
with concomitant normalized state
\begin{equation}
|\varphi_{-1,1}\rangle = \frac{1}{\sqrt{p(-1,1)}} 
\sum_{k \in S(-1,1)} \beta_{k}(t) | f_k\rangle
\end{equation}
and the set
\begin{equation}
S(-1,1)= \{13,17,21,25,28,30 \}.
\end{equation}
The result for the probability is shown in Figure (\ref{fig:41mp}).
The ancilla states can again be factored out
\begin{equation}
|\varphi_{-1,1}\rangle = |\tilde \varphi_{-1,1}\rangle \otimes
\keto \otimes \ketz
\end{equation}
but data qubits and resonators remain entangled.

The analogous results for obtaining the error syndrome $(1,-1)$ are
\begin{equation}
P_{1,-1} = \mathcal{I}_{r_1} \otimes \mathcal{I}_{r_2} \otimes
\mathcal{I}_{r_3} \otimes \mathcal{I}_{r_4}
 \otimes \mathcal{I}_{12} \otimes \ketz \braz \otimes \keto \brao,
\end{equation}
with probability
\begin{equation}
p(1,-1)=  \sum_{k \in S(1,-1)} \beta_k^*(t) \beta_k(t),
\end{equation}
state
\begin{equation}
|\varphi_{1,-1}\rangle = \frac{1}{\sqrt{p(1,-1)}} 
\sum_{k \in S(1,-1)} \beta_{k}(t) | f_k\rangle,
\end{equation}
and integer set
\begin{equation}
S(1,-1)= \{14,18,22,26,29,31 \}.
\end{equation}
After amputating the ancillas, the state reads
\begin{equation}
|\varphi_{1,-1}\rangle = |\tilde \varphi_{1,-1}\rangle \otimes
\ketz \otimes \keto.
\end{equation}
Since $p(1,-1) < 10^{-4}$, it is not shown.

The error syndrome $(-1,-1)$ corresponds to the operator
\begin{equation}
P_{-1,-1} = \mathcal{I}_{r_1} \otimes \mathcal{I}_{r_2} \otimes
\mathcal{I}_{r_3} \otimes \mathcal{I}_{r_4}
 \otimes \mathcal{I}_{12} \otimes \keto \brao \otimes \keto \brao,
\end{equation}
which projects to one basis state
\begin{equation}
|\varphi_{-1,-1}\rangle = \frac{\beta_{32}(t)}{|\beta_{32}(t)|}
| f_{32} \rangle,
\end{equation}
with probability
\begin{equation}
p(-1,-1)=   \beta_{32}^*(t) \beta_{32}(t).
\end{equation}
This probability is less than $10^{-7}$.
The final state is a product state in all degrees of freedom
\begin{equation}
|\varphi_{-1,-1}\rangle =
| 0 \rangle \otimes | 0 \rangle \otimes | 0 \rangle \otimes | 0 \rangle \otimes
\ketz \otimes \ketz \otimes \keto \otimes \keto,
\label{eq:state32}
\end{equation}
where we have omitted a phase factor.

\subsection{Bell state $|\Psi^+\rangle$}
\begin{figure}[htb]
\centering
\includegraphics[width=210pt]{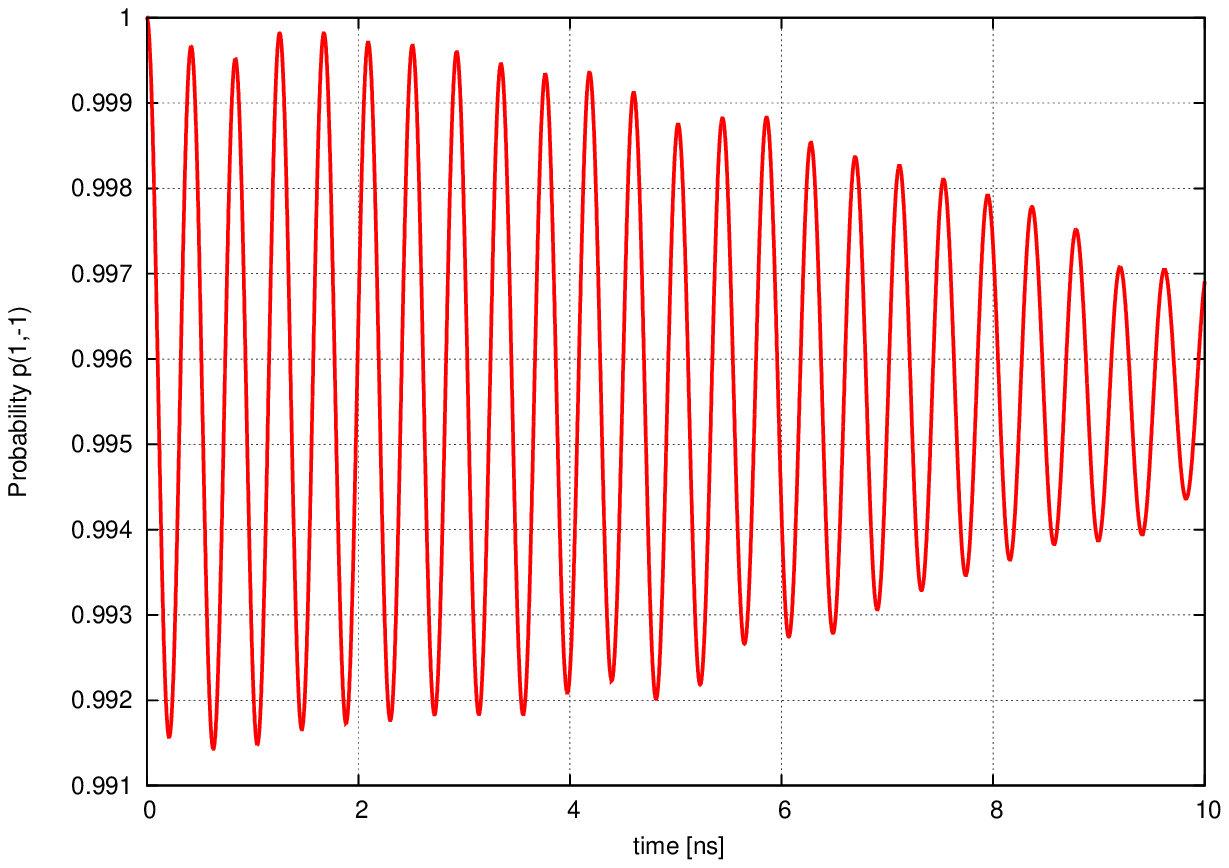}
\caption{\label{fig:42pm} Probability for obtaining the error syndrome (1,-1)
after unitary evolution of the state $|0,0,0,0\rangle |\Psi^+\rangle |\mathbf{0}, \mathbf{1}\rangle$
as a function of time.
Parameters as in Figure (\ref{fig:41pp}).}
\end{figure}

We do a similar analysis for the Bell state $|\Psi^+\rangle$: 
\begin{eqnarray}
|\psi_0\rangle &=& |0,0,0,0\rangle \otimes |\Psi^+\rangle \otimes 
|\mathbf{0}, \mathbf{1}\rangle  \nonumber \\
&=& \tfrac{1}{2}\sqrt{2}\left( |f_{29}\rangle + |f_{31}\rangle\right).
\label{eq:psi0}
\end{eqnarray}
Unitary evolution is governed by $U_2(t)$ and it yields
\begin{eqnarray}
|\psi(t) \rangle &=& 
\tfrac{1}{2}\sqrt{2}\, U_2(t)\left(|f_{29}\rangle + |f_{31}\rangle \right) \nonumber \\
&=& \tfrac{1}{2}\sqrt{2} \sum_{\eta=1}^{32}
e^{-iE_{2\eta}t} (b_{\eta 29}+b_{\eta 31}) |E_{2\eta}\rangle \nonumber \\
&=& \tfrac{1}{2}\sqrt{2} \sum_{\eta=1}^{32}
\sum_{k=1}^{32} e^{-iE_{2\eta}t} (b_{\eta 29}+b_{\eta 31}) b_{\eta k} | f_k\rangle \nonumber \\
&=& \sum_{k=1}^{32} \gamma_{k}(t) | f_k\rangle,
\end{eqnarray}
where we have abbreviated
\begin{equation}
\gamma_k(t) = \tfrac{1}{2}\sqrt{2} \sum_{\eta=1}^{32}
 e^{-iE_{2\eta}t} (b_{\eta 29}+b_{\eta 31}) b_{\eta k}.
\end{equation}
At this instant of time $t$ the measurement on the ancillas is done.
The measurements operators are defined above.
The error syndrome corresponding to no error detection reads $(1,-1)$.
We obtain for the probability for obtaining these values
\begin{equation}
p(1,-1)=  \sum_{k \in S(1,-1)} \gamma_k^*(t) \gamma_k(t),
\end{equation}
as shown in Figure (\ref{fig:42pm}).
In order not to overload the notation we do {\it not} introduce a
different symbol for probabilities in different examples.
The corresponding normalized state is 
\begin{equation}
|\psi_{1,-1}\rangle = \frac{1}{\sqrt{p(1,-1)}} 
\sum_{k \in S(1,-1)} \gamma_{k}(t) | f_k\rangle.
\end{equation}
The ancilla states can of course  be separated
\begin{equation}
|\psi_{1,-1}\rangle = |\tilde \psi_{1,-1}\rangle \otimes
\ketz \otimes \keto
\end{equation}
but once more the data qubits and resonators remain entangled.

\begin{figure}[htb]
\centering
\includegraphics[width=210pt]{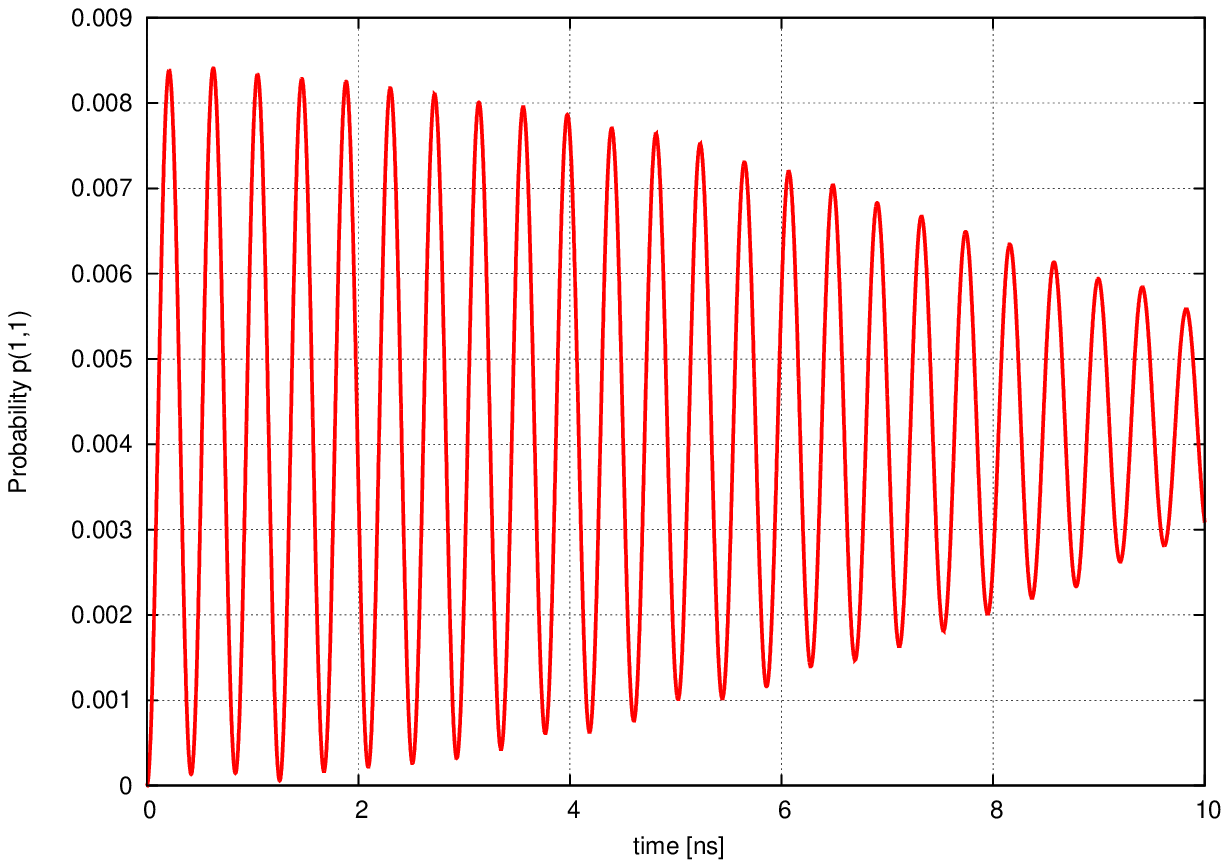}
\caption{\label{fig:42pp} Probability for obtaining the error syndrome (1,1)
after unitary evolution of the state $|0,0,0,0\rangle |\Psi^+\rangle |\mathbf{0}, \mathbf{1}\rangle$
as a function of time.
Parameters as in Figure (\ref{fig:41pp}).}
\end{figure}

The expressions for obtaining error syndromes indicating an error can be derived analogously
and we merely present these without further comments:
\begin{equation}
p(1,1)=  \sum_{k \in S(1,1)} \gamma_k^*(t) \gamma_k(t),
\end{equation}
\begin{eqnarray}
|\psi_{1,1}\rangle &=& \frac{1}{\sqrt{p(1,1)}} 
\sum_{k \in S(1,1)} \gamma_{k}(t) | f_k\rangle \nonumber \\
&=& |\tilde \psi_{1,1}\rangle \otimes
\ketz \otimes \ketz.
\end{eqnarray}
Figure (\ref{fig:42pp}) depicts the probability $p(1,1)$.

For the error syndrome $(-1,1)$ we get
\begin{equation}
p(-1,1)=  \sum_{k \in S(-1,1)} \gamma_k^*(t) \gamma_k(t),
\end{equation}
\begin{eqnarray}
|\psi_{-1,1}\rangle &=& \frac{1}{\sqrt{p(-1,1)}} 
\sum_{k \in S(-1,1)} \gamma_{k}(t) | f_k\rangle \nonumber \\
&=& |\tilde \psi_{-1,1}\rangle \otimes
\keto \otimes \ketz.
\end{eqnarray}
Because $p(-1,1) < 10^{-5}$ it is not shown.

Finally, we obtain for the syndrome $(-1,-1)$
\begin{equation}
p(-1,-1)=   \gamma_{32}^*(t) \gamma_{32}(t),
\end{equation}
which is less than $10^{-3}$. The final state
\begin{equation}
|\psi_{-1,-1}\rangle = \frac{1}{\sqrt{p(-1,-1)}} 
\gamma_{32}(t) | f_{32}\rangle
\end{equation}
coincides -up to a phase factor- with (\ref{eq:state32}).

\section{Fidelities and error correction}\label{sec:FIDEC}
In the last section, we have analyzed the four-qubit stabilizer system.
Unitary evolution of the complete coupled system initially containing 
a separable stable Bell state has been calculated.
There are two cases to be distinguished on the basis of the found error syndrome.
In the first one, the syndrome indicates no error is 
and hence no error correction is applied. Secondly, the syndrome indicates
a specific  error  which
subsequently may be corrected by applying the correction operators on the data qubits, cf. section (\ref{sec:EC}).
In both cases, one then is interested in a measure how well the stabilizer performs.
Note that in all cases the desired stable two-qubit state, or `target state', is known.
Thus a natural choice would be the fidelity of the realized state.
We have seen that in most cases, however, the two data qubits and the resonators
are entangled and the concomitant two-qubit state is not defined.
As a consequence, one needs to exploits the density matrix formalism \cite{Cohen1,NC} and the
technique of partial tracing. The latter reduces the density matrix of 
a composite system to the density matrix of a subsystem by performing the partial  trace
over the other degrees of freedom.
The reduced density matrix of the data qubits is therefore obtained by
tracing out the resonator degrees of freedom.
Note, however, in case of entanglement
the resulting subsystem density matrix will {\it not} correspond to a pure state.
In our examples, the resonators-qubits state is of course pure and, with the exception
of the product state after obtaing the error syndrome $(-1,-1)$, entangled.
Consequently, we cannot identify a pure two-qubit state since the density
matrix corresponds to a mixed state. Nevertheless, we can use
the concept of the fidelity between a pure (target) state $|\psi \rangle$
and the mixed state described by a density operator $\rho$. It is defined as
\begin{equation}
  F(|\psi\rangle,\rho)= \sqrt{\langle\psi|\rho|\psi\rangle}, 
  \label{eq:Fsd1}
\end{equation}
cf. \cite{NC,Carlo}.

\subsection{Density matrix: Bell state $|\Phi^+\rangle$}
\begin{figure}[htb]
\centering
\includegraphics[width=210pt]{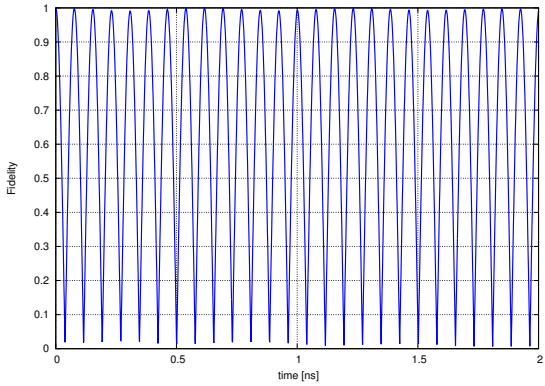}
\caption{\label{fig:F1pp} Fidelity
as a function of time
after unitary evolution of the state $|0,0,0,0\rangle |\Phi^+\rangle |\mathbf{0}, \mathbf{0}\rangle$
and subsequently obtaining error syndrome $(1,1)$.
Parameters as in Figure (\ref{fig:41pp}).}
\end{figure}

Once more we start the analysis for the case of the error syndrome $(1,1)$.
The state after having factored out the ancillas (\ref{eq:phipl11}) is
rewritten as
\begin{equation}
  |\tilde \phi_{1,1}\rangle= \frac{1}{\sqrt{p(1,1)}}
 \sum_{k \in \tilde S} \beta_k(t) 
| \tilde f_k \rangle,
\end{equation}
thereby (implicitly) defining $ | \tilde f_k \rangle$,
$ \beta_0(t)=\tfrac{1}{2}\sqrt{2}e^{-iE_0t}$
and $\tilde S= \{0\} \cup S(1,1)$.
The concomitant density matrix is therefore 
\begin{equation}
\rho =\frac{1}{p(1,1)} \sum_{k \in \tilde S} \sum_{l \in \tilde S}
\beta_k(t) \beta^*_l(t) |\tilde f_k\rangle \langle \tilde f_l|.
\label{eq:dmphi}
\end{equation}
At this point we take the partial trace with respect to the resonator 
degrees of freedom. The result for the reduced data qubit denstity
operator is given by
\begin{widetext}
\begin{eqnarray}
\label{eq:dm1}
\rho_q &=& \frac{1}{p(1,1)} \biggl\{ \sum_{k=0}^{10} \beta^*_k(t)\beta_k(t) 
| \mathbf{0,0} \rangle \langle\mathbf{0,0}|
+ \beta^*_{27}(t)\beta_{27}(t) | \mathbf{1,1} \rangle \langle\mathbf{1,1}|
\nonumber \\
&+& \sum_{k=11,15,19,23} \beta^*_k(t)\beta_k(t) 
| \mathbf{1,0} \rangle \langle\mathbf{1,0}|
+ \sum_{k=12,16,20,24} \beta^*_k(t)\beta_k(t) 
| \mathbf{0,1} \rangle \langle\mathbf{0,1}|
\nonumber \\
&+&  \beta_0(t)\beta^*_{27}(t) | \mathbf{0,0} \rangle \langle\mathbf{1,1}|
+  \beta_{27}(t)\beta^*_{0}(t) | \mathbf{1,1} \rangle \langle\mathbf{0,0}| \\
&+& \left(\beta_{11}(t)\beta^*_{12}(t) +\beta_{15}(t)\beta^*_{16}(t)
+\beta_{19}(t)\beta^*_{20}(t) +\beta_{23}(t)\beta^*_{24}(t)\right)
| \mathbf{1,0} \rangle \langle\mathbf{0,1}|
\nonumber \\ 
&+& \left(\beta_{12}(t)\beta^*_{11}(t) +\beta_{16}(t)\beta^*_{15}(t)
+\beta_{20}(t)\beta^*_{19}(t) +\beta_{24}(t)\beta^*_{23}(t)\right)
| \mathbf{0,1} \rangle \langle\mathbf{1,0}| \biggr\}.
\nonumber 
\end{eqnarray}
We eventually obtain the fidelity 
\begin{eqnarray}
\label{eq:F1pp}
F(|\Phi^+\rangle, \rho_q) = \sqrt{\langle \Phi^+ | \rho_q |\Phi^+\rangle}
&=& \left\{ \frac{1}{2p(1,1)}  \biggl[
\sum_{k=0}^{10} \beta^*_k(t) \beta_k(t)
 + \beta^*_{27}(t)\beta_{27}(t) + \beta^*_0(t) \beta_{27}(t) 
+ \beta^*_{27}(t) \beta_0(t)\biggr]\right\}^{1/2},
\nonumber
\end{eqnarray}
which is shown in Figure (\ref{fig:F1pp}).
\end{widetext}

After having detected an error, one may apply a correction operator to the state.
We analyze the consequences by computing
the fidelity of the obtained  target state and the reduced density
operator. Since the partial trace is taken with respect to the
resonators degrees of freedom and the error correction 
only operates on the data qubits, we can first compute the reduced density matrix and
eventually apply the error correction operator on the target state.
We demonstrate the method for the various error syndromes indicating an error.
 
\begin{figure}[htb]
\centering
\includegraphics[width=210pt]{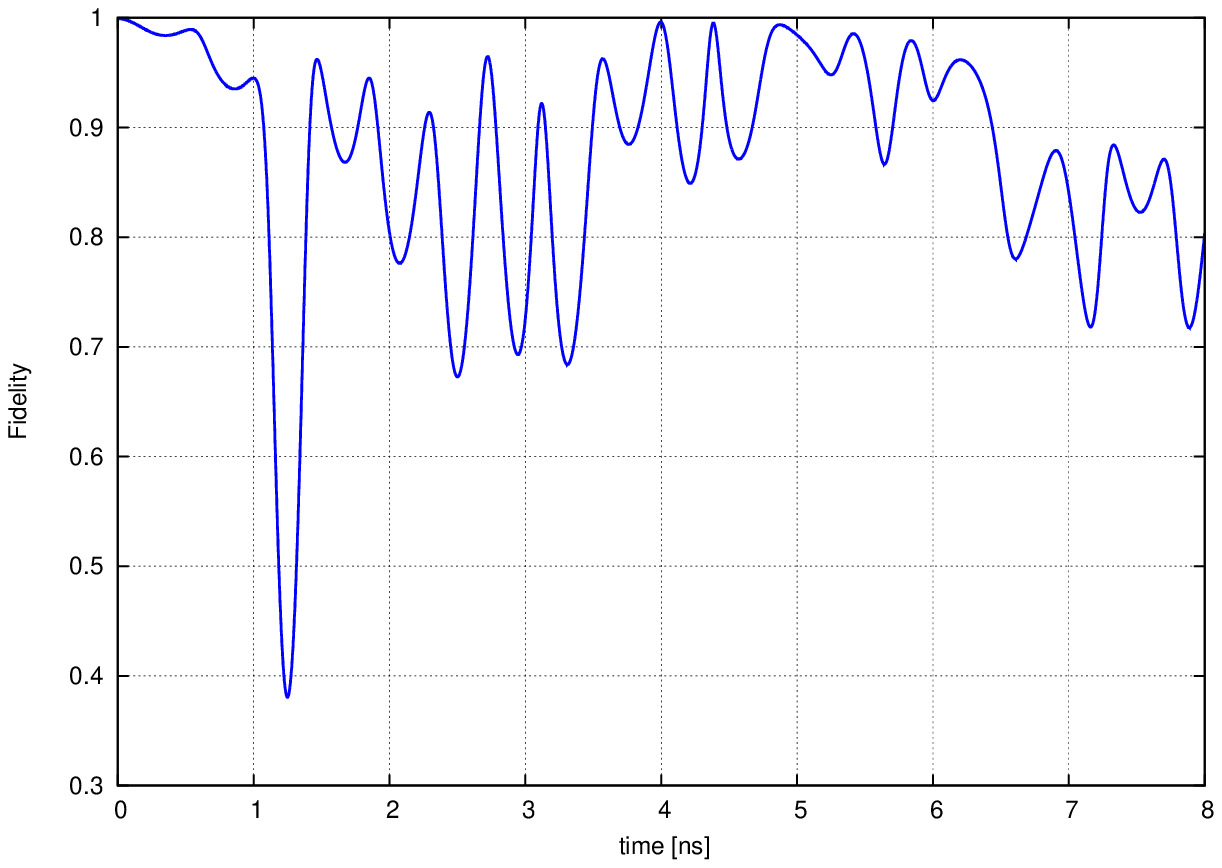}
\caption{\label{fig:F1pm} Fidelity
as a function of time
after unitary evolution of the state $|0,0,0,0\rangle |\Phi^+\rangle |\mathbf{0}, \mathbf{0}\rangle$,
subsequently obtaining error syndrome $(1,-1)$ and error correction.
Parameters as in Figure (\ref{fig:41pp}).}
\end{figure}

For the syndrome  $(-1,1)$ the state after factoring out the ancillas reads
\begin{equation}
  |\tilde \phi_{-1,1}\rangle= \frac{1}{\sqrt{p(-1,1)}}
 \sum_{k \in S(-1,1)} \beta_k(t) 
| \tilde f_k \rangle,
\end{equation}
with corresponding density matrix
\begin{equation}
\rho =\frac{1}{p(-1,1)} \sum_{k \in S(-1,1)} \sum_{l \in  S(-1,1)}
\beta_k(t) \beta^*_l(t) |\tilde f_k\rangle \langle \tilde f_l|.
\end{equation}
The reduced qubit density matrix follows as
\begin{eqnarray}
\label{eq:dm2}
\rho_q &=& \frac{1}{p(-1,1)}\biggl\{ 
\sum_{k=13,17,21,25} \beta^*_k(t)\beta_k(t) 
| \mathbf{0,0} \rangle \langle\mathbf{0,0}| \\
&+& \beta_{28}(t)\beta^*_{28}(t) | \mathbf{1,0} \rangle \langle\mathbf{1,0}|
+\beta_{30}(t)\beta^*_{30}(t) | \mathbf{0,1} \rangle \langle\mathbf{0,1}|
\nonumber \\
&+& \beta_{28}(t)\beta^*_{30}(t) | \mathbf{1,0} \rangle \langle\mathbf{0,1}|
+ \beta_{30}(t)\beta^*_{28}(t) | \mathbf{0,1} \rangle \langle\mathbf{1,0}| \biggr\}. \nonumber
\end{eqnarray}
Herewith we get as fidelity
\begin{widetext}
\begin{eqnarray}
&& F(| \Phi^+\rangle, Z^{[1]} \rho_q Z^{[1]}) =
 \sqrt{\langle \Phi^+ |  Z^{[1]}\rho_q Z^{[1]}|\Phi^+\rangle}=
\sqrt{\langle \Phi^- |  \rho_q |\Phi^-\rangle} \nonumber\\
&=& \left\{ \frac{1}{2p(-1,1)}  \biggl[
 \beta^*_{13}(t) \beta_{13}(t)
 + \beta^*_{17}(t)\beta_{17}(t) + \beta^*_{21}(t) \beta_{21}(t) 
+ \beta^*_{25}(t) \beta_{25}(t)\biggr]\right\}^{1/2}, 
\end{eqnarray}
which is close to zero.

Analogously, for the error syndrome $(1,-1)$ the state after factoring out the ancillas is given by
\begin{equation}
  |\tilde \phi_{1,-1}\rangle= \frac{1}{\sqrt{p(1,-1)}}
 \sum_{k \in S(1,-1)} \beta_k(t) 
| \tilde f_k \rangle,
\end{equation}
with concomitant density matrix
\begin{equation}
\rho =\frac{1}{p(1,-1)} \sum_{k \in S(1,-1)} \sum_{l \in S(1,-1)}
\beta_k(t) \beta^*_l(t) |\tilde f_k\rangle \langle \tilde f_l|.
\end{equation}
Calculating the partial trace yields
\begin{eqnarray}
\label{eq:dm3}
\rho_q &=& \frac{1}{p(1,-1)}\biggl\{ 
\sum_{k=14,18,22,26} \beta^*_k(t)\beta_k(t) 
| \mathbf{0,0} \rangle \langle\mathbf{0,0}|
+ \beta_{29}(t)\beta^*_{29}(t) | \mathbf{1,0} \rangle \langle\mathbf{1,0}|
+\beta_{31}(t)\beta^*_{31}(t) | \mathbf{0,1} \rangle \langle\mathbf{0,1}|
\nonumber \\
&+& \beta_{29}(t)\beta^*_{31}(t) | \mathbf{1,0} \rangle \langle\mathbf{0,1}|
+ \beta_{31}(t)\beta^*_{29}(t) | \mathbf{0,1} \rangle \langle\mathbf{1,0}| \biggr\}.
\end{eqnarray}
Subsequently, we obtain the fidelity
\begin{eqnarray}
&& F(| \Phi^+\rangle, X^{[2]} \rho_q X^{[2]}) =
 \sqrt{\langle \Phi^+ |  X^{[2]}\rho_q X^{[2]}|\Phi^+\rangle}=
\sqrt{\langle \Psi^+ |  \rho_q |\Psi^+\rangle}\nonumber \\
&=& \left\{ \frac{1}{2p(1,-1)}  \biggl[
 \beta^*_{29}(t) \beta_{29}(t)
 + \beta^*_{31}(t)\beta_{31}(t) + \beta^*_{31}(t) \beta_{29}(t) 
+ \beta^*_{29}(t) \beta_{31}(t)\biggr]\right\}^{1/2}, 
\label{eq:F1pm}
\end{eqnarray}
as shown in Figure (\ref{fig:F1pm}).
\end{widetext}

In case of error syndrome $(-1,-1)$, it is easily seen that the
reduced density matrix corresponds to a the pure state 
\begin{equation}
 \rho_q = | \mathbf{0,0} \rangle \langle\mathbf{0,0}|.
\label{eq:dm4}
\end{equation} 
Then we get
\begin{eqnarray}
&& F(| \Phi^+\rangle, X^{[2]} Z^{[1]} \rho_q  Z^{[1]}X^{[2]}) \nonumber \\ &=&
 \sqrt{\langle \Phi^+ |  Z^{[1]} X^{[2]}\rho_q Z^{[1]} X^{[2]}|\Phi^+\rangle} \nonumber \\ &=&
\sqrt{\langle \Psi^- |  \rho_q |\Psi^-\rangle} =  0, 
\end{eqnarray}
that is a vanishing fidelity.

\begin{figure}[htb]
\centering
\includegraphics[width=210pt]{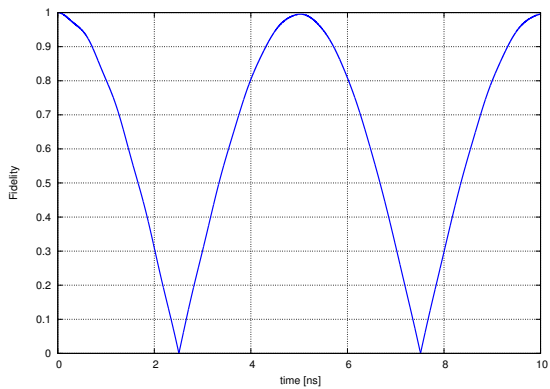}
\caption{\label{fig:F2pm} Fidelity
as a function of time
after unitary evolution of the state $|0,0,0,0\rangle |\Psi^+\rangle |\mathbf{0}, \mathbf{1}\rangle$
and subsequently obtaining error syndrome $(1,-1)$.
Parameters as in Figure (\ref{fig:41pp}).}
\end{figure}

\begin{widetext}
\subsection{Density matrix: Bell state $|\Psi^+\rangle$}
The computations for the Bell state 
  $|\Psi^+\rangle$ of course proceed completely analogously and we merely
present the results for the obtained fidelities.
For the error syndrome  $(1,-1)$, we get
\begin{equation}
F(| \Psi^+\rangle, \rho_q ) =
 \sqrt{\langle \Psi^+ |  \rho_q |\Psi^+\rangle} 
= \left\{ \frac{1}{2p(1,-1)}  \biggl[
 \gamma^*_{29}(t) \gamma_{29}(t)
 + \gamma^*_{31}(t)\gamma_{31}(t) + \gamma^*_{31}(t) \gamma_{29}(t) 
+ \gamma^*_{29}(t) \gamma_{31}(t)\biggr]\right\}^{1/2}.
\label{eq:F2pm}
\end{equation}
This fidelity is shown in Figure (\ref{fig:F2pm}).

In case of obtaining the error syndrome $(1,1)$, the resulting error corrected
fidelity follows as
\begin{equation}
F(|\Psi^+\rangle, X^{[1]} \rho_qX^{[1]})= 
\sqrt{\langle \Psi^+ |X^{[1]} \rho_q X^{[1]} |\Psi^+\rangle}=
\sqrt{\langle \Phi^+ | \rho_q |\Phi^+\rangle} 
= \left\{ \frac{1}{2p(1,1)}  \biggl[
\sum_{k=1}^{10} \gamma^*_k(t) \gamma_k(t)
 + \gamma^*_{27}(t)\gamma_{27}(t) \biggr]\right\}^{1/2},
\end{equation}
which is less than 0.0035 and therefore not shown here.

After error detection by means of the syndrome $(-1,1)$, we obtain as fidelity
\begin{eqnarray}
&& F(| \Psi^+\rangle, X^{[2]} Z^{[1]} \rho_q Z^{[1]}X^{[2]} ) =
 \sqrt{\langle \Psi^+ |  X^{[2]} Z^{[1]}\rho_q Z^{[1]}X^{[2]} |\Psi^+\rangle}=
\sqrt{\langle \Phi^- |  \rho_q |\Phi^-\rangle} \nonumber\\
&=& \left\{ \frac{1}{2p(-1,1)}  \biggl[
 \gamma^*_{13}(t) \gamma_{13}(t)
 + \gamma^*_{17}(t)\gamma_{17}(t) + \gamma^*_{21}(t) \gamma_{21}(t) 
+ \gamma^*_{25}(t) \gamma_{25}(t)\biggr]\right\}^{1/2}.
\end{eqnarray}
\end{widetext}
It turns out be aprroximately constant, very close to the value $\tfrac{1}{2}\sqrt{2}$.

The case with error syndrome $(-1,-1)$ once more yields a vanishing fidelity.

\section{Re-insertion into the circuit}\label{sec:RINS}
The stabilizer circuit has the property that re-inserting the stable state into
the circuit, reproduces it. The interesting question arises how this is
possibly influenced
by the unitary evolution governed by the Hamiltonian before the re-insertion. 
We analyze this problem once more for the two stable Bell states situations for
which we have already calculated the evolution. Note that we assume a certain
duration of the evolution before entering the circuit again. We 
emphasize that this duration is arbitrary but then fixed.

\subsection{Bell state $|\Phi^+\rangle$}
In this case, we have to consider as initial state
\begin{equation}
|\varphi_{\text{in}}\rangle = \frac{1}{\sqrt{p(1,1)}} 
\sum_{k \in \tilde S} \beta_{k}(t) | f_k\rangle.
\end{equation}
In \cite{Fowler} and appendix \ref{app:appB}, the circuit analysis has been presented. 
After expressing two-qubit computational basis states in terms of the Bell states
it can immediately be applied. The states of the resonators are not affected;
for these we use the short-hand notation $|R_k\rangle$. We obtain as output
state from the stabilizing circuit
\begin{eqnarray}
|\varphi_{\text{out}}\rangle &=& \frac{1}{\sqrt{2p(1,1)}} \biggl[
 \biggl\{ \sum_{k\in \bar S} \beta_k(t) |R_k\rangle 
\biggr\} \otimes |\Psi^+\rangle  \otimes|\mathbf{0, 1}\rangle \nonumber \\
 &+&
\biggl\{ \sum_{k=0}^{10} \beta_k(t) |R_k\rangle + \beta_{27}(t) |R_{27}\rangle
\biggr\} \otimes |\Phi^+\rangle  \otimes|\mathbf{0, 0}\rangle \nonumber \\
&+&  \biggl\{
\sum_{k=0}^{10} \beta_k(t) |R_k\rangle - \beta_{27}(t) |R_{27}\rangle
\biggr\} \otimes |\Phi^-\rangle  \otimes|\mathbf{1, 0}\rangle \nonumber \\
 &+& \biggl\{
\sum_{k\in \bar S} (-1)^k \beta_k(t) |R_k\rangle 
\biggr\} \otimes |\Psi^-\rangle  \otimes|\mathbf{1, 1}\rangle\biggr],
\end{eqnarray}
with the set $\bar S =\{11,12,15,16,19,20,23,24\}$.
Next we assume the ancillas to be measured and compute the probabilities
for the various error syndromes. Recall that for the stable state, these
are given by $p(1,1)=1; p(1,-1)=p(-1,1)=p(-1,-1)=0$.
Instead of these, we get probabilities $\tilde p$:
\begin{eqnarray}
\tilde p(1,1) &=&  \frac{1}{2p(1,1)}  \biggl[
\sum_{k=0}^{10} \beta^*_k(t) \beta_k(t)
 + \beta^*_{27}(t)\beta_{27}(t) \nonumber \\ &+& \beta^*_0(t) \beta_{27}(t) 
+ \beta^*_{27}(t) \beta_0(t)\biggr], \nonumber \\
\tilde p(1,-1) &=&  \frac{1}{2p(1,1)}  \biggl[
\sum_{k \in \bar S} \beta^*_k(t) \beta_k(t)  
 + \beta^*_{11}(t)\beta_{12}(t) \nonumber \\ &+& \beta^*_{12}(t) \beta_{11}(t) 
+ \beta^*_{15}(t) \beta_{16}(t) + \beta^*_{16}(t) \beta_{15}(t) \nonumber \\
 &+& \beta^*_{19}(t)\beta_{20}(t) + \beta^*_{20}(t) \beta_{19}(t) \nonumber \\
&+& \beta^*_{23}(t) \beta_{24}(t) + \beta^*_{24}(t) \beta_{23}(t) \biggr], \nonumber\\
\tilde p(-1,1) &=&  \frac{1}{2p(1,1)}  \biggl[
\sum_{k=0}^{10} \beta^*_k(t) \beta_k(t) \nonumber \\
 &+& \beta^*_{27}(t)\beta_{27}(t) - \beta^*_0(t) \beta_{27}(t) 
- \beta^*_{27}(t) \beta_0(t)\biggr], \nonumber \\
\tilde p(-1,-1) &=&  \frac{1}{2p(1,1)}  \biggl[
\sum_{k \in \bar S} \beta^*_k(t) \beta_k(t) 
 - \beta^*_{11}(t)\beta_{12}(t) \nonumber \\ &-& \beta^*_{12}(t) \beta_{11}(t) 
- \beta^*_{15}(t) \beta_{16}(t) - \beta^*_{16}(t) \beta_{15}(t) \nonumber \\
 &-& \beta^*_{19}(t)\beta_{20}(t) - \beta^*_{20}(t) \beta_{19}(t) \nonumber \\
&-& \beta^*_{23}(t) \beta_{24}( t) - \beta^*_{24}(t) \beta_{23}(t) \biggr]  .
\end{eqnarray}
It can be readily verified that they add up to one.
Comparison with (\ref{eq:F1pp})
yields the rather intriguing result
\begin{equation}
\tilde p(1,1) = F^2(|\Phi^+ \rangle, \rho_q),
\label{eq:intr}
\end{equation}
connecting the fidelity in terms of a reduced density matrix
to a measurable probability.
The probabilities $\tilde p(1,-1), \tilde p(-1,-1)$ are of order $10^{-9}$;
consequently $p(-1,1) \simeq 1-p(1,1)$.

\subsection{Bell state $|\Psi^+\rangle$}
Once more, we repeat the calculation for the Bell state $|\Psi^+\rangle$.
The state after unitary evolution reads in this case
\begin{equation}
|\psi_{1,-1}\rangle = \frac{1}{\sqrt{p(1,-1)}} 
\sum_{k \in S(1,-1)} \gamma_{k}(t) | f_k\rangle.
\end{equation}
The input state for the stabilizer follows from re-initializing
the ancillas in the ground state 
\begin{equation}
|\psi_{\text{in}}\rangle = \frac{1}{\sqrt{p(1,-1)}} 
\sum_{k \in S(1,-1)} \gamma_{k}(t) | \tilde f_k\rangle \otimes |\mathbf{0,0}\rangle.
\end{equation}
The next steps of the stabilizing circuit produce
\begin{eqnarray}
|\psi_{\text{out}}\rangle &=& \frac{1}{\sqrt{2p(1,-1)}} \biggl[ \biggl\{
\sum_{k=14,18,22,26} \gamma_k(t) |R_k\rangle
\biggr\} \nonumber \\ &\otimes& |\Phi^+\rangle  \otimes|\mathbf{0, 0}\rangle \\
 &+&  \biggl\{ \gamma_{29}(t) |R_{29}\rangle 
 + \gamma_{31}(t) |R_{31}\rangle 
\biggr\} \otimes |\Psi^+\rangle  \otimes|\mathbf{0, 1}\rangle \nonumber \\
&+&  \biggl\{
\sum_{k=14,18,22,26} \gamma_k(t) |R_k\rangle
\biggr\} \otimes |\Phi^-\rangle  \otimes|\mathbf{1, 0}\rangle \nonumber \\
 &+&  \biggl\{ \gamma_{31}(t) |R_{31}\rangle 
 - \gamma_{29}(t) |R_{29}\rangle 
\biggr\} \otimes |\Psi^-\rangle  \otimes|\mathbf{1, 1}\rangle \biggr]. \nonumber
\end{eqnarray}
The corresponding probabilities follow as
\begin{eqnarray}
\tilde p(1,-1) &=& \frac{1}{2p(1,-1)}\bigl[
 \gamma^*_{29}(t) \gamma_{29}(t)
 +\gamma^*_{31}(t) \gamma_{31}(t) \nonumber \\
 &+&\gamma^*_{29}(t) \gamma_{31}(t)
 +\gamma^*_{31}(t) \gamma_{29}(t)\bigr], \nonumber \\
\tilde p(1,1) &=& \frac{1}{2p(1,-1)}
\sum_{k=14,18,22,26} \gamma_k^*(t) \gamma_k(t), \nonumber \\
\tilde p(-1,1) &=& \tilde p(1,1), \\
\tilde p(-1,-1) &=& \frac{1}{2p(1,-1)}\bigl[
 \gamma^*_{29}(t) \gamma_{29}(t)
 +\gamma^*_{31}(t) \gamma_{31}(t) \nonumber \\
 &-&\gamma^*_{29}(t) \gamma_{31}(t)
 -\gamma^*_{31}(t) \gamma_{29}(t)\bigr]. \nonumber 
\end{eqnarray}
As is easily checked, the sum of these probabilities equals one. 
Once more, we get a relation like (\ref{eq:intr})
\begin{equation}
\tilde p(1,-1) = F^2(|\Psi^+ \rangle, \rho_q).
\label{eq:fid-prob1}
\end{equation}
Both probabilities $p(1,-1)$ and $\tilde p(1,1)$ are of order $10^{-9}$.
Therefore, we get $\tilde p(-1,-1) \simeq 1-\tilde p(1,-1)$.

\subsection{Fidelity and re-insertion}
The equality of the considered re-insertion probabilities and the squares
of the fidelities (\ref{eq:intr}, \ref{eq:fid-prob1}) is not coincidental.
We will prove and generalize this result.
For the two examples of stable Bell states, we write the  complete
state after unitary evolution as
\begin{equation}
|\varphi(t)\rangle = \sum\limits_{k=0}^{32} \mu_k(t) |f_k\rangle,
\end{equation}
where for Bell state $|\Phi^+\rangle, \mu_k(t)=\beta_k(t) \, (k\ge 0)$
and for Bell state $|\Psi^+\rangle, \mu_0(t)=0, \mu_k(t)=\gamma_k(t) \, (k\ge 1)$.
The ancilla measurement operators for obtaining $m_{a,b} = \pm  1$,
are written as
\begin{equation}
P_{m_a,m_b} = \mathcal{I}_{r_1} \otimes \mathcal{I}_{r_2} \otimes
\mathcal{I}_{r_3} \otimes \mathcal{I}_{r_4}
 \otimes \mathcal{I}_{12} \otimes |\mathbf{n_a}\rangle \langle \mathbf{n_a} |
 \otimes 
 |\mathbf{n_b}\rangle \langle \mathbf{n_b}|,
\end{equation}
with $\mathbf{n_a, n_b} \in \left\{\mathbf{0,1}\right\}$. Recall that the measurement results (error syndromes) $1,-1$
 correspond  respectively to the ancilla states $\ketz, \keto$.
Operating on the state and subsequent normalization yields 
\begin{eqnarray}
&&P_{m_a,m_b} |\varphi(t)\rangle \rightarrow
|\tilde \varphi(t)\rangle = \\ && \frac{1}{\sqrt{p(m_a,m_b)}} \sum\limits_{k\in S(m_a,m_b)} \mu_k(t) |q_k\rangle |R_k\rangle
\otimes |\mathbf{n_a, n_b} \rangle, \nonumber
\end{eqnarray}
with measurement probability $p(m_a,m_b)$ and data two-qubit state $|q_k\rangle$.
The set $S(m_a,m_b)$ labels states corresponding with the obtained error syndrome. 
After this measurement, the ancillas can be amputated resulting in
\begin{equation}
|\bar \varphi(t)\rangle = \frac{1}{\sqrt{p(m_a,m_b)}} \sum\limits_{k \in S(m_a,m_b)} \mu_k(t) |q_k\rangle |R_k\rangle.
\label{eq:ampstate}
\end{equation}
This state is the starting point for the computation of respective fidelities or to analyze the re-insertion into
the stabilizer circuit.

The latter starts with adding the re-initialized ancillas and use as circuit input
\begin{equation}
|\psi(t)\rangle = \frac{1}{\sqrt{p(m_a,m_b)}} \sum\limits_{k \in S(m_a,m_b)} \mu_k(t) |q_k\rangle |R_k\rangle \ketz \ketz.
\end{equation}
The data qubit states are expanded as
\begin{equation}
|q_k\rangle
 = \alpha_k |\Phi^+\rangle + \beta_k |\Psi^+\rangle + \gamma_k |\Phi^-\rangle + \zeta_k |\Psi^-\rangle
\label{eq:expa}
\end{equation}
with normalization $ |\alpha_k|^2 + |\beta_k|^2 + |\gamma_k|^2 + |\zeta_k|^2 = 1$,
cf. (\ref{eq:genstat}). The allegedly ideal circuit is linear and we
assume it does not affect the photon modes in the resonators.
According to \cite{Fowler} and appendix \ref{app:appB}, its output state is then given by
\begin{widetext}
\begin{eqnarray}
|\Psi\rangle &=& \frac{1}{\sqrt{p(m_a,m_b)}} \sum\limits_{k \in S(m_a,m_b)} \mu_k(t) |R_k\rangle \\ &\otimes& \left(
  \alpha_k |\Phi^+\rangle \ketz \ketz + \beta_k |\Psi^+\rangle  \ketz \keto+ \gamma_k |\Phi^-\rangle \keto \ketz + \zeta_k |\Psi^-\rangle \keto \keto\right),
\nonumber
\end{eqnarray}
yielding the following ancilla measurement probabilities which, although not explicitly indicated, depend on $m_a, m_b$
\begin{eqnarray}
\tilde p(1,1) &=& \frac{1}{p(m_a,m_b)} \sum\limits_{k \in S(m_a,m_b)} \sum\limits_{l \in S(m_a,m_b)}
\alpha_k \alpha_l^*
\mu_k(t) \mu_l^*(t) \langle R_l |R_k\rangle, \nonumber \\
\tilde p(1,-1) &=& \frac{1}{p(m_a,m_b)} \sum\limits_{k \in S(m_a,m_b)} \sum\limits_{l \in S(m_a,m_b)}
\beta_k \beta_l^* \mu_k(t) \mu_l^*(t) \langle R_l |R_k\rangle, \nonumber \\
\tilde p(-1,1) &=& \frac{1}{p(m_a,m_b)} \sum\limits_{k \in S(m_a,m_b)} \sum\limits_{l \in S(m_a,m_b)}
\gamma_k \gamma_l^* \mu_k(t) \mu_l^*(t) \langle R_l |R_k\rangle, \nonumber \\
\tilde p(-1,-1) &=& \frac{1}{p(m_a,m_b)} \sum\limits_{k \in S(m_a,m_b)} \sum\limits_{l \in S(m_a,m_b)}
\zeta_k \zeta_l^* \mu_k(t) \mu_l^*(t) \langle R_l |R_k\rangle. 
\label{eq:probf}
\end{eqnarray}
Note that $\langle R_l | R_k \rangle $ may also be equal to one for $k \ne l$ which lead to the earlier
obtained interference terms.

The state after amputating the ancillas $|\bar \varphi(t)\rangle$,
cf.  (\ref{eq:ampstate}),
corresponds to the density matrix
\begin{equation}
\bar \rho = \frac{1}{p(m_a,m_b)} 
\sum\limits_{k \in S(m_a,m_b)} \sum\limits_{l \in S(m_a,m_b)}
\mu_k(t) \mu_l^*(t) |q_k\rangle |R_k\rangle \langle q_l |\langle R_l |.
\end{equation}
We explicitly introduce a complete set of resonator states as $|r_n \rangle$ and take
the partial trace with respect to these degrees of freedom. In this way we obtain the two-qubit
density matrix
\begin{eqnarray}
\rho_q &=& \frac{1}{p(m_a,m_b)} \sum\limits_{k \in S(m_a,m_b)} \sum\limits_{l \in S(m_a,m_b)} \sum\limits_n
\mu_k(t) \mu_l^*(t) \langle r_n| R_k\rangle \langle R_l | r_n\rangle |q_k\rangle \langle q_l|, \nonumber \\
 &=& \frac{1}{p(m_a,m_b)} \sum\limits_{k \in S(m_a,m_b)} \sum\limits_{l \in S(m_a,m_b)}
\mu_k(t) \mu_l^*(t) \langle R_l |R_k\rangle |q_k\rangle \langle q_l|.
\end{eqnarray}
\end{widetext}
Using the expansion (\ref{eq:expa}) yields the fidelities
\begin{eqnarray}
F^2(|\Phi^+\rangle,\rho_q) &=& \langle \Phi^+| \rho_q | \Phi^+ \rangle = \tilde p(1,1), \nonumber \\
F^2(|\Psi^+\rangle,\rho_q) &=& \langle \Psi^+| \rho_q | \Psi^+ \rangle = \tilde p(1,-1), \nonumber \\
F^2(|\Phi^-\rangle,\rho_q) &=& \langle \Phi^-| \rho_q | \Phi^- \rangle = \tilde p(-1,1), \nonumber \\
F^2(|\Psi^-\rangle,\rho_q) &=& \langle \Psi^-| \rho_q | \Psi^- \rangle = \tilde p(-1,-1).
\end{eqnarray}
Hence we have established, for a given error syndrome $m_a,m_b$, identities between squared fidelities and measurement
probabilities after re-insertion of the respective states into the ideal stabilizer circuit.

\section{Free evolution \& Rotating frame}\label{sec:FRF}
It may be instructive to repeat some of the computations for the noninteracting system.
Of course, the free system also unitarily evolves in time. 
In order to partially separate consequences of free evolution a transformation to 
a rotating frame is useful.
Formally a rotating frame transformation is equivalent to adopting the interaction
representation, see e.g. \cite{Merz}.

\subsection{Evolution without interaction}
The Hamiltonian $H_0$ is obtained by setting all couplings equal to zero and
corresponds to free data qubits, free ancillas and uncoupled resonators.
We start with the initial state defined in eq.(\ref{eq:ini}), rewritten as
\begin{equation}
|\varphi_0\rangle =\tfrac{1}{2}\sqrt{2} |0,0,0,0\rangle \otimes  
(|\mathbf{0}, \mathbf{0}\rangle + |\mathbf{1}, \mathbf{1}\rangle)
\otimes |\mathbf{0}, \mathbf{0}\rangle.
\end{equation}
It is a superposition of the ground state and of an excited state
of the `free' Hamiltonian. We denote the latter by $|\tilde E \rangle$
with corresponding energy $\tilde E= E_0+\omega_1'+\omega_2'$.
Unitary evolution governed by $H_0$ yields 
\begin{equation}
|\varphi(t) \rangle =\tfrac{1}{2}\sqrt{2} e^{-iE_0t} 
\left( |E_0\rangle  + e^{-i\omega_+t} |\tilde E\rangle\right), 
\label{eq:psipfree}
\end{equation}
where $\omega_+=\omega_1'+\omega_2'$.
It follows that measuring the ancillas gives with probability one the error syndrome $(1,1)$.
So far, so good; note the difference with the interacting case.
Since there is no entanglement in this noninteracting problem,
the reduced density operator still corresponds  to a pure state
\begin{equation}
|\varphi_q(t) \rangle =\tfrac{1}{2}\sqrt{2}  
\left( |\mathbf{0,0} \rangle  + e^{-i\omega_+t} |\mathbf{1,1} \rangle\right). 
\end{equation}
Nevertheless, the obtained fidelity is oscillatory in the elapsed evolution time
\begin{equation}
F=|\langle \Phi^+|\varphi_q(t) \rangle| =\tfrac{1}{2}\sqrt{2} \sqrt{1+\cos{\omega_+t}}.
\label{eq:Ffreep}
\end{equation}
Our earlier results indeed coincide in the limit of vanishing couplings.

As above, we can re-insert the state after evolution (\ref{eq:psipfree}) into the stabilizer.
Changing the basis gives the following input state for fixed elapsed evolution time $t$:
\begin{eqnarray}
|\varphi_{\text{in}} \rangle &=& \tfrac{1}{2} e^{-iE_0 t} |0,0,0,0\rangle \otimes
\left\{\left(1+e^{-iw_+ t}\right)|\Phi^+\rangle \right.\nonumber \\
&+&\left.\left(1-e^{-iw_+ t}\right)|\Phi^-\rangle\right\} \otimes|\mathbf{0,0} \rangle.
\end{eqnarray}
The stabilizer then yields as output state
\begin{eqnarray}
|\varphi_{\text{out}} \rangle &=&  \tfrac{1}{2} e^{-iE_0 t} \bigl[  
\left(1+e^{-iw_+ t}\right)
|0,0,0,0\rangle \otimes |\Phi^+\rangle \otimes|\mathbf{0,0} \rangle \nonumber\\
&+&\left(1-e^{-iw_+  t}\right)
|0,0,0,0\rangle \otimes |\Phi^-\rangle \otimes|\mathbf{1,0} \rangle \bigr] .
\end{eqnarray}
The ancilla measurement probabilities follow as
\begin{eqnarray}
\tilde p(1,1) &=&  \tfrac{1}{2}\left(1+\cos{\omega_+ t}\right), \nonumber \\
\tilde p(-1,1) &=&  \tfrac{1}{2}\left(1-\cos{\omega_+ t}\right), \nonumber \\
\tilde p(1,-1) &=& \tilde p(-1,-1) =  0.
\end{eqnarray}
The sum of  the probabilities is obviously one and the fidelity found in (\ref{eq:Ffreep})
satisfies
\begin{equation}
F^2 = \tilde p(1,1).
\end{equation}

Similar results are obtained if we start with the initial state
involving the Bell state $|\Psi^+ \rangle$, that is eq.(\ref{eq:psi0}).
Unitary evolution according to $H_0$ leads to
\begin{eqnarray}
|\psi(t) \rangle &=& \tfrac{1}{2}\sqrt{2} e^{-i(E_0+\omega_1'+\omega_b)t} |0,0,0,0\rangle \nonumber \\ &\otimes&
\left\{ |\mathbf{1,0}\rangle + e^{i\omega_-t} |\mathbf{0,1}\rangle
\right\} \otimes|\mathbf{0,1} \rangle,
\end {eqnarray}
with $\omega_-=\omega_1'-\omega_2'$.
It can readily be checked that measuring the $Z$ operators of the ancillas yields
$(1,-1)$ with probability one. Recall the different results for the interacting system. 
The fidelity, however, again shows oscillations in the elapsed time $t$
\begin{equation}
F=|\langle \Psi^+|\varphi_q(t) \rangle| =\tfrac{1}{2}\sqrt{2} \sqrt{1+\cos{\omega_- t}}.
\label{eq:Ffreeps}
\end{equation}
After re-initialization, the corresponding input state for the stabilizer is written as
\begin{eqnarray}
|\psi_{\text{in}} \rangle &=&\tfrac{1}{2} e^{-i(E_0+\omega_1'+\omega_b)t} |0,0,0,0\rangle \\  &\otimes&
\left\{\left(1+e^{iw_- t}\right)|\Psi^+\rangle
+\left(e^{iw_- t}-1\right)|\Psi^-\rangle\right\} \otimes|\mathbf{0,0} \rangle. \nonumber
\end {eqnarray}
The circuit then yields as output the state 
\begin{eqnarray}
|\psi_{\text{out}} \rangle &=&  \tfrac{1}{2} e^{-i(E_0+\omega_1'+\omega_b) t} |0,0,0,0\rangle \nonumber  \\&\otimes&
\bigl[  \left(1+e^{iw_- t}\right)
|\Psi^+\rangle \otimes|\mathbf{0,1} \rangle \nonumber\\
&+&\left(e^{iw_+ t}-1\right)
\otimes |\Psi^-\rangle \otimes|\mathbf{1,1} \rangle \bigr] .
\end{eqnarray}
It leads to the following probabilities for the error syndromes
\begin{eqnarray}
\tilde p(1,-1) &=&  \tfrac{1}{2}\left(1+\cos{\omega_-t}\right), \nonumber \\
\tilde p(-1,-1) &=&  \tfrac{1}{2}\left(1-\cos{\omega_- t}\right), \nonumber \\
\tilde p(1,1) &=& \tilde p(-1,1) =  0.
\end{eqnarray}
The probabilities add up to one and again the relation
between the fidelity (\ref{eq:Ffreeps}) and probability $\tilde p$
\begin{equation}
F^2 = \tilde p(1,-1),
\end{equation}
is confirmed.

\subsection{Rotating frame}

\subsubsection{Data qubits}
The rotating frame of the two data qubits is most
important. The evolution due to the Hamiltonian
\begin{equation}
H_0^{[12]} = H_0^{[1]} + H_0^{[2]} =
-\tfrac{1}{2} \omega_1' \sigma_z^{[1]} -\tfrac{1}{2} \omega_2' \sigma_z^{[2]},
\end{equation}
is effectively taken into account.
 We explicitly perform the transformation
\begin{equation}
R(t) = \exp{[iH_0^{[12]}t]} = \exp{[iH_0^{[1]}t]} \exp{[iH_0^{[2]}t]},
\end{equation}
where the identity holds because $[H_0^{[1]}, H_0^{[2]}]=0$.
The analysis of section \ref{sec:UEBM} is shortly
repeated  for the rotating frame. To this end, we note that 
the introduced basis states are eigenstates of the operator $R(t)$
\begin{equation}
R(t) |f_k\rangle  = \exp{[iH_0^{[12]}t]} |f_k\rangle =
e^{i(\epsilon_k^{[1]}+\epsilon_k^{[2]})t} |f_k\rangle,
\end{equation}
where $\epsilon_k^{[1]}= \pm \tfrac{1}{2}\omega_1',
\epsilon_k^{[2]}= \pm \tfrac{1}{2}\omega_2'$, depending on the index $k$.
In the following we will explicitly need
\begin{eqnarray}
&&\epsilon_0^{[1]}+ \epsilon_0^{[2]} = -\tfrac{1}{2}\omega_+, \quad
\epsilon_{27}^{[1]}+ \epsilon_{27}^{[2]} = \tfrac{1}{2}\omega_+, \nonumber \\
&&\epsilon_{29}^{[1]}+ \epsilon_{29}^{[2]} = \tfrac{1}{2}\omega_-, \quad
\epsilon_{31}^{[1]}+ \epsilon_{31}^{[2]} = -\tfrac{1}{2}\omega_- .
\end{eqnarray}
It is also convenient to define the modified functions
\begin{equation}
\tilde \beta_k(t) = e^{i(\epsilon_k^{[1]}+\epsilon_k^{[2]})t} \beta_k(t), \quad
\tilde \gamma_k(t) = e^{i(\epsilon_k^{[1]}+\epsilon_k^{[2]})t} \gamma_k(t).
\end{equation}
Now we proceed to the evolution of two previously analyzed Bell states.\\

\begin{figure}[htb]
\centering
\includegraphics[width=210pt]{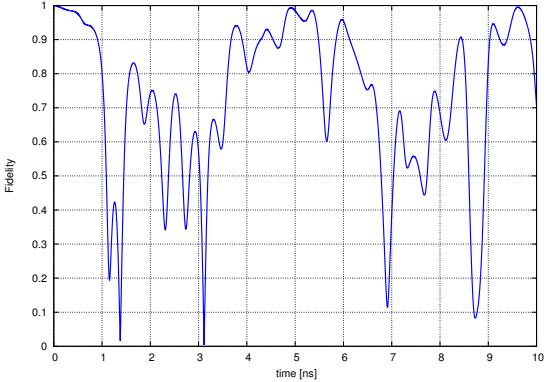}
\caption{\label{fig:F1pmrf} Fidelity in the rotating frame
as a function of time
after unitary evolution of the state $|0,0,0,0\rangle |\Phi^+\rangle |\mathbf{0}, \mathbf{0}\rangle$,
subsequently obtaining error syndrome  $(1,-1)$ and error correction.
Parameters as in Figure (\ref{fig:41pp}). Compare to Figure (\ref{fig:F1pm}).}
\end{figure}

\begin{figure}[htb]
\centering
\includegraphics[width=210pt]{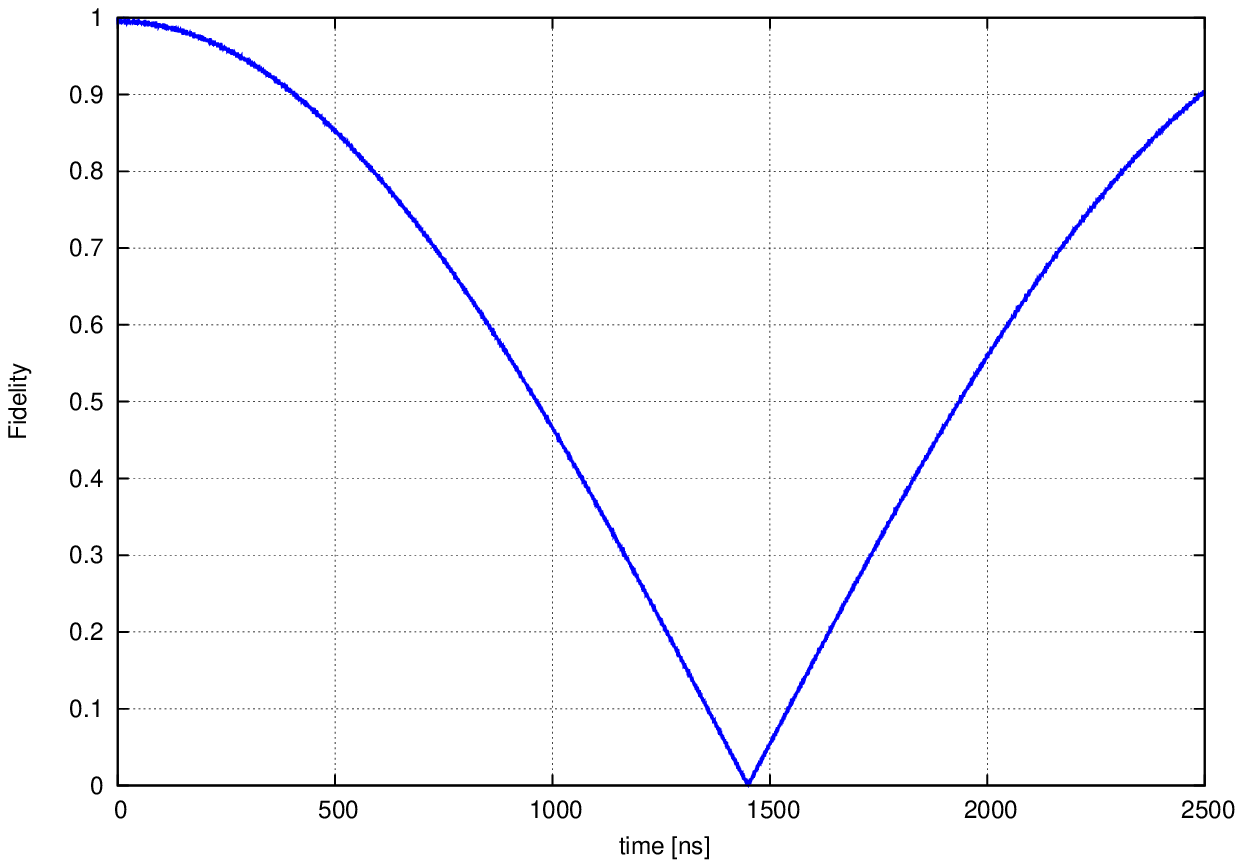}
\caption{\label{fig:F2pmrf} Fidelity in the rotating frame
as a function of time
after unitary evolution of the state $|0,0,0,0\rangle |\Psi^+\rangle |\mathbf{0}, \mathbf{0}\rangle$
and subsequently obtaining error syndrome result $(1,-1)$.
Parameters as in Figure (\ref{fig:41pp}). Compare to Figure (\ref{fig:F2pm}).}
\end{figure}

\paragraph{Bell state $|\Phi^+\rangle$\\}
The evolved state (\ref{eq:euphi}) in the rotated frame is given by
\begin{equation}
|\phi_R(t)\rangle = R(t) |\phi(t) \rangle = \sum_{k=0}^{32}\tilde \beta_k(t) |f_k\rangle.
\end{equation}
Since $\tilde \beta(t) \tilde \beta^*(t) = \beta(t) \beta^*(t)$, the measurement probabilities
of obtaining the various error syndromes do not change. 
The density matrix (\ref{eq:dmphi}) corresponding
to the syndrome $(1,1)$, however, does change:
\begin{equation}
\rho^R =\frac{1}{p(1,1)} \sum_{k \in \tilde S} \sum_{l \in \tilde S}
\tilde \beta_k(t) \tilde \beta^*_l(t) |\tilde f_k\rangle \langle \tilde f_l|.
\end{equation}
As a consequence, the fidelity (\ref{eq:F1pp}) is also modified 
\begin{figure}[htb]
\centering
\includegraphics[width=210pt]{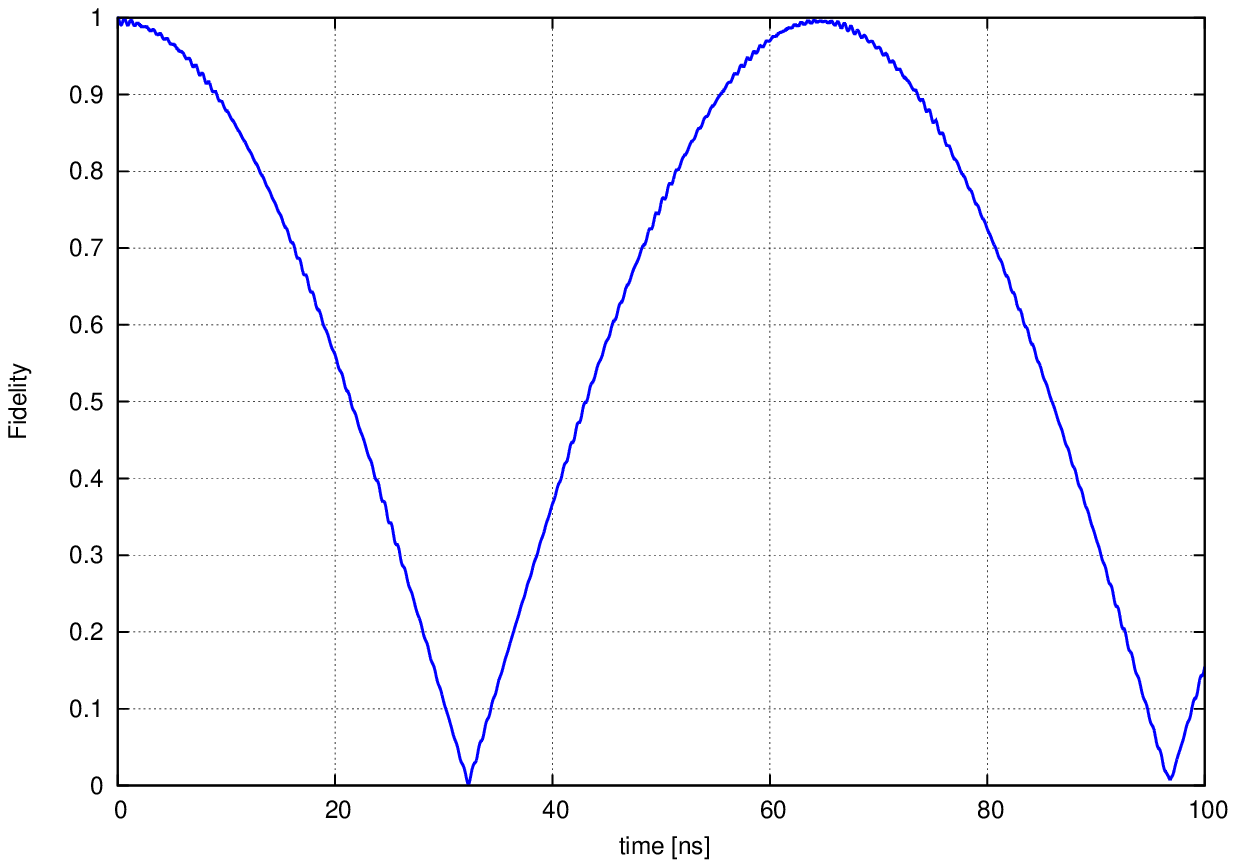}
\caption{\label{fig:F1pprf} Fidelity in the rotating frame
as a function of time
after unitary evolution of the state $|0,0,0,0\rangle |\Phi^+\rangle |\mathbf{0}, \mathbf{0}\rangle$
and subsequently obtaining error syndrome $(1,1)$.
Parameters as in Figure (\ref{fig:41pp}). Compare to Figure (\ref{fig:F1pp}).}
\end{figure}
\begin{widetext}
\begin{eqnarray}
&&F_R(|\Phi^+\rangle, \rho^R_q) =  \sqrt{\langle \Phi^+ | \rho^R_q |\Phi^+\rangle}\\
&=& \left\{ \frac{1}{2p(1,1)}  \biggl[
\sum_{k=0}^{10} \beta^*_k(t) \beta_k(t)
 + \beta^*_{27}(t)\beta_{27}(t) + \tilde \beta^*_0(t) \tilde \beta_{27}(t) 
+ \tilde \beta^*_{27}(t) \tilde \beta_0(t)\biggr]\right\}^{1/2} \nonumber \\
&=& \left\{ \frac{1}{2p(1,1)}  \biggl[
\sum_{k=0}^{10} \beta^*_k(t) \beta_k(t)
 + \beta^*_{27}(t)\beta_{27}(t) + e^{i\omega_+t} \beta^*_0(t) \beta_{27}(t) 
+ e^{-i\omega_+t}  \beta^*_{27}(t)  \beta_0(t)\biggr]\right\}^{1/2}, \nonumber
\end{eqnarray}
as shown in Figure (\ref{fig:F1pprf}). It is readily verified that the relation between
probalility for finding result $(1,1)$ 
after re-insertion into the stabilizer and fidelity 
(\ref{eq:intr}) is also valid in the rotating frame:
\begin{equation}
\tilde p_R(1,1) = F^2(|\Phi^+ \rangle, \rho^R_q).
\end{equation}

Analogous computations for the error syndrome $(1,-1)$ yield
a modification of the fidelity (\ref{eq:F1pm}) as well
\begin{eqnarray}
&& F_R(| \Phi^+\rangle, X^{[2]} \rho^R_q X^{[2]}) =
 \sqrt{\langle \Phi^+ |  X^{[2]}\rho^R_q X^{[2]}|\Phi^+\rangle}=
\sqrt{\langle \Psi^+ |  \rho_q |\Psi^+\rangle}\\
&=& \left\{ \frac{1}{2p(1,-1)}  \biggl[
 \beta^*_{29}(t) \beta_{29}(t)
 + \beta^*_{31}(t)\beta_{31}(t) + \tilde \beta^*_{31}(t)  \tilde\beta_{29}(t) 
+  \tilde\beta^*_{29}(t)  \tilde\beta_{31}(t)\biggr]\right\}^{1/2} \nonumber \\
&=& \left\{ \frac{1}{2p(1,-1)}  \biggl[
\beta^*_{29}(t) \beta_{29}(t)
 +  \beta^*_{31}(t)\beta_{31}(t) +  e^{i\omega_-t}\beta^*_{31}(t) \beta_{29}(t) 
+ e^{-i\omega_-t} \beta^*_{29}(t) \beta_{31}(t)\biggr]\right\}^{1/2}. \nonumber
\end{eqnarray}
This result is depicted in Figure (\ref{fig:F1pmrf}).
The final fidelities after having obtained the error syndromes $(-1,1)$ and $(-1,-1)$
do not change in the rotating frame. The reason is that these expressions do
not contain interference between $\beta_k(t)$ functions with different indices $k$.\\

\paragraph{Bell state $|\Psi^+\rangle\\$}
The unitarily evolved state corresponding to the two qubit Bell state $|\Psi^+\rangle$
reads in the rotating frame
\begin{equation}
|\psi_R(t)\rangle = R(t) |\psi(t) \rangle = \sum_{k=1}^{32}\tilde \gamma_k(t) |f_k\rangle.
\end{equation}
The measurement probabilities once more do not change because
$\tilde \gamma(t) \tilde \gamma^*(t) = \gamma(t) \gamma^*(t)$. 
The density matrix resulting from the error syndrome $(1,-1)$ 
has two interference terms and, as a consequence, it gets  modified. Concomitantly,
we obtain a modification of the fidelity (\ref{eq:F2pm}): 
\begin{eqnarray}
&&F_R(| \Psi^+\rangle, \rho^R_q ) =
 \sqrt{\langle \Psi^+ |  \rho^R_q |\Psi^+\rangle} \\
&=& \sqrt{ \frac{1}{2p(1,-1)}  \biggl[
 \gamma^*_{29}(t) \gamma_{29}(t)
 + \gamma^*_{31}(t)\gamma_{31}(t) + \tilde \gamma^*_{31}(t)  \tilde\gamma_{29}(t) 
+  \tilde\gamma^*_{29}(t)  \tilde\gamma_{31}(t)\biggr]} \nonumber\\
&=& \sqrt{ \frac{1}{2p(1,-1)}  \biggl[
 \gamma^*_{29}(t) \gamma_{29}(t)
 + \gamma^*_{31}(t)\gamma_{31}(t) + e^{i\omega_-t} \gamma^*_{31}(t) \gamma_{29}(t) 
+  e^{-i\omega_-t}\gamma^*_{29}(t) \gamma_{31}(t)\biggr]}. \nonumber
\end{eqnarray}
\end{widetext}
This result is shown in Figure (\ref{fig:F2pmrf}).
The relation between probability after re-insertion and fidelity is once again valid 
\begin{equation}
\tilde p_R(1,-1) = F^2(|\Psi^+ \rangle, \rho^R_q).
\end{equation}
The other fidelities do not contain interference terms and are therefore not 
changed in the rotating frame.

\subsubsection{Including ancillas}
It is straightforward to extend the rotating frame with the ancillas.
Their `free' Hamiltonians commute and they also commute with  $H_0^{[12]}$.
If we denote the extended operator as $\mathcal R$, we obtain
\begin{equation}
\mathcal R(t) |f_k\rangle  = 
e^{i(\epsilon_k^{[1]}+\epsilon_k^{[2]}+\epsilon_k^{[a]}+\epsilon_k^{[b]})t} |f_k\rangle,
\end{equation}
with $ \epsilon_k^{[a]}= \pm \tfrac{1}{2}\omega_a, \epsilon_k^{[b]}= \pm \tfrac{1}{2}\omega_b$,
once more  depending on the index $k$.
The relevant energies are given by
\begin{eqnarray}
&&\epsilon_0^{[1]}+ \epsilon_0^{[2]} =
-\tfrac{1}{2}\omega_+ -\tfrac{1}{2}\omega_a -\tfrac{1}{2}\omega_b, \nonumber\\
&&\epsilon_{27}^{[1]}+ \epsilon_{27}^{[2]} = \tfrac{1}{2}\omega_-
-\tfrac{1}{2}\omega_a -\tfrac{1}{2}\omega_b , \nonumber \\
&&\epsilon_{29}^{[1]}+ \epsilon_{29}^{[2]} = \tfrac{1}{2}\omega_-
-\tfrac{1}{2}\omega_a +\tfrac{1}{2}\omega_b , \nonumber\\
&&\epsilon_{31}^{[1]}+ \epsilon_{31}^{[2]} = -\tfrac{1}{2}\omega_- 
-\tfrac{1}{2}\omega_a +\tfrac{1}{2}\omega_b.
\end{eqnarray}
The functions $\beta_k(t)$ and $\gamma_k(t)$ are now modified as
\begin{eqnarray}
\hat \beta_k(t) &=& e^{i( \epsilon_k^{[1]}+\epsilon_k^{[2]} +\epsilon_k^{[a]}+\epsilon_k^{[b]})t} \beta_k(t),
\nonumber \\
\hat \gamma_k(t) &=& e^{i( \epsilon_k^{[1]}+\epsilon_k^{[2]} +\epsilon_k^{[a]}+\epsilon_k^{[b]})t} \gamma_k(t).
\end{eqnarray}
The calculation now proceeds as above but now with the functions 
$\hat \beta_k(t)$ and $\hat \gamma_k(t)$. The new phase factors due to the ancillas, however,
cancel in the terms which could have altered the previous results for the fidelities.
As a consequence, the obtained fidelities do {\it not} change in this extended rotated frame. 

\section{Theoretical discussion}\label{sec:THEO}
\subsection{Error analysis}
The stabilizing circuit with two ancilla qubits and two data qubits is a building block
in surface codes for error correction \cite{Fowler,Brav,Terhal}. Of course, this has motivated our choice for
these investigations. The circuit, including the ancilla measurement and possibly error correction
can deal with errors on the data qubits which occur before performing the circuit. 
The unitary evolution we have analyzed is supposed to take place between the circuit
unitary operations and ancilla measurements. Even if one would be able to interpret the consequences
as `errors', it is shown that they cannot be corrected in the usual way, that is applying
the usual transformation on the data qubits based on the error syndromes.  

With respect to the interpretation as `errors', there is no obvious identification of the
error type.  
In \cite{Devitt} coherent gate errors,
environmental decoherence, initialization and measurement errors as well as loss and leakage
are discussed. One may consider the electromagnetic resonators and even the ancillas
as `environment'; the ancillas are measured, the remaining resonators are traced out.
It will be further analyzed below using the concept of quantum operations \cite{NC,Caves}.
At this point,  we can already state that it does not yield a satisfactory correction mechanism.
Our findings are more similar to those in \cite{gate-error}, where a gate-error analysis
for transmon qubits is presented. The unitary evolution of the total system is also calculated in that study.
As in our case, it is inherently unitary but yields detrimental effects on gates
in the computational subspace. The `errors' are also systematic but appear  
incoherent in that subspace. In both analyses, they originate from entanglement
with other states due to Hamiltonian evolution; 
in \cite{gate-error} noncomputational transmon states and the resonator --
in our work, ancillas and resonators.

Although no quantum code for the stabilizing circuit is developed as such, there is a certain resemblance because of the
use of certain subspaces in a larger Hilbert space. Quantum error correction by means of coding,
e.g. Shor's nine-bit code, is for instance explained in \cite{Caves}. Within a complete Hilbert space
a code subspace is defined. The considered error operators map the code subspace to orthogonal
subspaces. Errors can then be detected, distinguished and eventually corrected by determining in which
subspace the system is. It can be done without disturbance and destroying the coherence; superposition and
entanglement are essential in the process. Returning to the stabilizing circuit, we recall that a
product state of a general two-data qubit state and the ancillas in the ground state is put in. 
The initial subspace is therefore four-dimensional, whereas the complete qubit Hilbert space is sixteen-dimensional. 
After the defined operations, the final state also belongs to a  four-dimensional subspace, having
the specific structure given in appendix \ref{app:appB}. The ancilla measurements subsequently project to one
of the one-dimensional subspaces of the data qubit Hilbert space.
However, if the final state evolves unitarily  according to the system Hamiltonian then the four-dimensional
subspace is abandoned. It explains why this particular error identification and correction does not work.
Assume that the initial state is one of the Bell states. Without evolution the corresponding probabilities
for the measurement outcomes are zero or one. This is no longer true after unitary evolution. Two more aspects
are important.
First, even if the ancilla measurement indicates no error, the final data qubit state is not the expected one.
Secondly, if the syndrome indicates an error, the usual error correction transformations,
cf. section \ref{sec:EC}, do not work.
Note that these findings do not depend merely on the inclusion of the resonators as degrees of freedom.
Indeed, in a model with only spin-exchange interaction between adjacent qubits, similar results have
been obtained \cite{spinstab}. Of course, in the model with resonators one has to extend
the ideal qubit state with resonators; here we have taken a product state with empty resonators.
There remains a noticeable
difference in the results compared to the spin model.
Since a final data qubit state has to be identified the resonators have to be traced out
after the evolution;  it does not matter whether this is done before or after the measurement, see \ref{sub:quan}.
In any case, the final data qubit `state' is not a pure state -as in the exchange model- but can only be described by means
of a density operator. In both models, however, the essential phenomenon is that the evolution operator
leads the state out of the stable subspace.

\subsection{Stabilizers in the Heisenberg picture}
The concept of the $X, Z$ stabilizing circuit has also been demonstrated by calculating with stabilizers instead
of states \cite{Fowler}. To this end, the Heisenberg representation is used. In particular,
it is necessary to perform CNOT operations on products of the $X$, $Z$ and $ \mathcal I$ operators.
Here we do not reproduce the derivation of appendix B of \cite{Fowler}, but only note that 
the resulting two stabilizers are $X^{[a]} X^{[1]} X^{[2]} \mathcal I^{[b]}$
and  $\mathcal I^{[a]} Z^{[1]} Z^{[2]} Z^{[b]}$. These operators commute and 
ancilla measurements indeed yield the two-qubit stabilizers $Z^{[1]} Z^{[2]}$  and
$X^{[1]} X^{[2]}$. However, in case of unitary evolution before the measurement
we need to  evolve the four-qubit stabilizers since we are in the Heisenberg representation.
Without doing the explicit calculation, it is easily seen that the two stabilizers are not invariant
under time-evolution. Recalling that in the Heisenberg equation for an operator
$A$ the commutator $[A,H]$ drives non-trivial evolution, the commutators of the two four-qubit stabilizers
with the full Hamiltonian are to be calculated. Here we only note that these commutators are nonzero
causing the stabilizers to change. Equivalently, this can be seen using the explicitly constructed
evolution operator $U(t)$ to find the evolution $A(t)= U(t) A U^\dagger(t)$. Once again, without doing the computation
completely, it is clear that the stabilizers evolve non-trivially. As a consequence, the stabilizing
mechanism will fail after finite time $t$. Resonator operators will appear in the evolution
of the stabilizers. After the evolution time $t$ they need to be traced out gain.
The latter does of course not apply to the interacting spin system \cite{spinstab}.
Nevertheless, the stabilizers also evolve in time, reflecting the degradation of the stabilizer formalism
taking into account full Hamiltonian evolution. 

In larger systems on which a surface code can be implemented similar observations hold.
Consider, for example,  Surface 17 \cite{Versluis}, which contains the studied two data qubits, two ancillas 
sytem as subsystems. Surface 17 consists out of nine data qubits and eight ancillas. In this system,
eight stabilizers can be identified \cite{Fowler}. These constrain the $2^9$-dimensional data qubit space to
a large extent.
However, a two-dimensional subspace is not constrained. It can be exploited as computational space
spanned by two logical qubits $|{\bf 0_L}\rangle, |{\bf 1_L}\rangle$.
Concomitantly, logical operators $Z_L, X_L$ can be defined. Our concerns based on unitary evolution
describing the complete dynamics apply here as well. Stabilizers and logical operators evolve in time
and the associated subspaces are not invariant. Recently, a numerical study testing quantum fault tolerance
on small systems has appeared \cite{Will}.

\subsection{Quantum operations}\label{sub:quan}
The framework of quantum operations has been developed to deal with quantum noise by means of  error correction
\cite{NC, Caves}. The eventual error correction is typically done for the code subspace which, of course,
is defined for a specific code. For example the three qubit bit flip code uses the two-dimensional
code subspace spanned by the states $|000 \rangle$ and $|111\rangle$.   
The formalism aims at taking into account the interaction of the system of interest and the environment.
It includes selective dynamics, where the environment is not observed, as well
as nonselective dynamics, where a measurement on the environment is made.
Various approaches for formulating quantum operations are presented in \cite{NC}.

Here we use quantum operations to describe the processes in our system. Note that we do not consider a
quantum code and, consequently, there is no code subspace. Noise is not considered and, concomitantly,
there is no actual environment.
 We may, however, consider or rather describe the  electromagnetic resonators
as environment since we eventually focus on the qubits by taking the partial trace with respect to
these degrees of freedom. Because the resonators are not measured, this dynamics is nonselective.
In addition, the ancillas can be considered as environment as well. The reason is that they are
measured also implying that their dynamics is selective. We will see below that the order of
tracing out resonators and measuring the ancillas is not relevant.

Quantum operations are defined as operators transforming density matrices.
For the full system we merely use the symbol $\rho$,
for data qubits $\rho_q$, for ancillas $\rho_a$ and for resonators $\rho_r$. The ideal circuit
has the transformation property
\begin{equation}
|b_m\rangle
|\,\mathbf{0,0}\rangle
\langle b_m|
\langle \mathbf{0,0}\,| \;
\longrightarrow \;
|b_m\rangle
|a_m\rangle
\langle a_m|
\langle b_m |, 
\end{equation}
where $|b_m\rangle, m\in\{1,2,3,4\}$ denotes one of the Bell states with corresponding
ancilla state $| a_m \rangle$. Next we extend this matrix with empty resonators thereby defining
the initial density matrix before unitary evolution
\begin{eqnarray}
\rho(0) &=& 
|0, 0, 0, 0\rangle
|b_m\rangle
|a_m\rangle
\langle a_m|
\langle b_m|
\langle 0, 0, 0, 0 | \nonumber \\
&=& \rho_r(0) \otimes \rho_q(0) \otimes \rho_a(0).
\end{eqnarray}
Unitary evolution yields the density matrix
\begin{equation}
\rho(t) =  U(t)\left(
 \rho_r(0) \otimes \rho_q(0) \otimes \rho_a(0)\right) U^\dagger(t).
\end{equation}
Note that we explicitly have constructed the evolution operator up to (and including) the
second excitation level. Measuring the ancillas is decribed by the operators  $P_k$ where
$k=(1,1) \dots (-1,-1)$ labels the measurement result, {\it i.e.}, the error syndrome. 
The subsequent `amputation'
of the projected ancilla state can formally be done via a partial trace. Finally we trace out
the resonators to obtain the quantum operation 
\begin{eqnarray}
\rho_q(0) &\rightarrow& \rho_q(t)=  \mathcal E(\rho_q)
 \\ &=&
\text{tr}_r\left[ \text{tr}_a\left[ P_k U(t) \left(
 \rho_r(0) \otimes \rho_q(0) \otimes \rho_a(0)\right) U^\dagger P_k \right]\right]. \nonumber
\end{eqnarray}
Alternatively, we  take the partial trace with respect to the resonators before the measurement
\begin{eqnarray}
\tilde \rho_q(0) &\rightarrow& \tilde \rho_q(t)=  \tilde{\mathcal E}(\rho_q) \\ &=&
\text{tr}_a\left[ P_k \, \text{tr}_r\left[ U(t) \left(
 \rho_r(0) \otimes \rho_q(0) \otimes \rho_a(0)\right) U^\dagger\right] P_k \right].\nonumber
\end{eqnarray}
Inserting the initial density matrix yields
\begin{equation}
 \mathcal E(\rho_q)= \tilde{\mathcal E}(\rho_q)= \sum_r A_{kr} \rho_q A^\dagger_{kr},
\end{equation}
with the data qubit operators
\begin{equation}
 A_{kr} = \langle r  a_k | U(t) |r_0 a_m\rangle. 
\end{equation}
The resonator states are here denoted by $|r\rangle$ and $|r_0 \rangle = | 0,0,0,0 \rangle$.
The result generalizes the simplest operator-sum representation \cite{NC}, because of the additional trace 
over the resonators. Its normalized form is given by
\begin{equation}
 \mathcal E(\rho_q)= \frac{\sum_r A_{kr} \rho_q A^\dagger_{kr}}
{\text{tr}\left[\sum_r A_{kr} \rho_q A^\dagger_{kr}\right]}, 
\label{eq:fin-den}
\end{equation}
where the normalization equals the probability that the measurement outcome labelled $k$
is obtained for the defined initial Bell state. The operators $A$, often denoted by $E$, are called operational
elements in \cite{NC}; in our case they depend on the evolution time.
The theory of quantum error correction exploits these operators to derive conditions for
error correction on a subspace defined by a certain quantum code \cite{NC}. Note that in our case
error correction would mean reversing the unitary evolution for the data qubits. Since this appears
to be unfeasible in practice, we do not pursue this.  

We conclude this section by noting that we have already derived explicit expressions
for the quantum operations (\ref{eq:fin-den}). For the initial Bell state $|\Phi^+\rangle$
they are given in (\ref{eq:dm1}), (\ref{eq:dm2}), (\ref{eq:dm3}) and (\ref{eq:dm4}).
The trivial, {\it i.e.}, time-independent pure computational basis state,
 result (\ref{eq:dm4}) can be easily `corrected' because
\begin{equation}
C_N \, \left(\bm{H}^{[1]} \otimes \mathcal I^{[2]}\right)|\,\mathbf{0,0}\rangle
=|\Phi^+\rangle,
\end{equation}
where $\bm{H}$ is the Hadamard transformation and $C_N$ denotes the CNOT operation.

\subsection{Synopsis}
At this point, it may be useful to summarize the results of our study.
First, we have computed unitary evolutions of the extended output states
of the stabilized circuit. The extension is to include the empty resonators. 
For free evolution, {\it i.e.}, zero couplings, the error syndrome indicating no error
is obtained with certainty. This does not longer hold for the interacting
case, where these probabilities become somewhat smaller than one. Concomitantly,
the probabilities for a different syndrome which  indicates an error are nonzero.
The measurement operator for the error syndrome indicating no  error does in general {\em not}
project the evolved state on the desired target two-qubit resonator state. 
The exception is the case where the initial state is an eigenstate
of the Hamiltonian which governs the time evolution.
There are two related consequences. Calculating the fidelity of the two-qubit
subsystem with the target Bell states, which is done by taking the partial trace
of the resonator degrees of freedom, yields oscillating fidelities
with high frequencies. Secondly, re-inserting the evolved two-qubit resonator stated
completed with (possibly) re-initialized ancillas in the ground state, does not lead
to the desired target state. The concomitant measurement probabilities
resonate once again with high frequencies.

Similar results follow for error-corrected two-qubit states constructed after the first
error detection. Neither high fidelities nor subsequent desired measurement probabilities
close to one are obtained. It happens for free as well as complete, that is including the interactions,
evolution. In other words, Hamiltonian governed unitary evolution causes
problems for the stabilizing circuit as the target states are no energy
eigenstates. For free evolution, however, target states are linear superpositions
of eigenstates and, as a consequence, the probability of getting the  syndrome indicating no error
remains one.  In the interacting case, even the latter breaks down 
although the mentioned probability is close to one whereas the error detection
probabilities are rather small. Another difference for nonzero coupling is the
fact that the two-qubit subsystem can only be described with a density
matrix which does {\em not} correspond to a pure state. This is a consequence of the
entanglement of data qubits and the resonators, even after the measurement
of the ancillas. For the noninteracting system, the two-qubit resonator system
is described by a product state and taking the partial trace yields a density
matrix which describes a pure state. In fact, the results for $H_0$ can be obtained
without invoking the density matrix formalism.

The calculations have also been done in the rotating frame of the data qubits. Measurement
probabilities for the various error syndromes are not altered. The fidelities, however,
do change. The observed time variations are slower, which indicates the expected cancellation
of the highest frequency components. Nevertheless,  with one exception, the resulting fidelities   
decrease rather fast. Extending the rotating frame with respect to the ancillas does
not change the results.

\section{Conclusion}\label{sec:CONC}
Our semi-analytical analysis of the $X$ and $Z$ stabilizing circuit has revealed
deteriorating consequences
of unitary evolution of the complete, {\it i.e.}, qubits and resonators, interacting system.
Probabilities for error syndromes indicating no error, therefore ideally equal one,
decrease to the range $0.992-0.999$.
Fidelities of the augmented density matrices and the stable target state, also ideally one,
oscillate between $0$ and $1$.
After having obtained an error syndrome indicating an error, the usual error correction
does not apply. The latter is {\it 
a forteriori} explained because unitary evolution does lead out of subspaces on which the usual
error correction does work.  The explanation in terms of the stabilizers of the circuits is that
they are no longer conserved, {\it i.e.}, evolve non-trivially governed by the complete Hamiltonian.

Several caveats should be kept in mind with respect to this research.
We have only addressed part of the effects of unitary evolution, that is
only in a time interval between the last unitary operation in the stabilizing  circuit
and the ancilla measurement. The latter is supposed to be perfect. It is also
 assumed that the circuit performs perfectly.
The rotating wave approximation has been made in the interaction terms.  Environmental
decoherence is neglected. Only the four-qubit system has been considered,  etc. 

Nevertheless, we think that
the results can be relevant for the further development of quantum error correction and
fault-tolerant computation. Recall that these concepts rely on the
applicability of some mathematical error model, like one-qubit
Pauli channels \cite{NC}. The latter cannot describe the deleterious effects
on the computational subspace we have identified.
The preliminary, somewhat discomforting, overall conclusion is that
the existing stabilizer codes are not sufficient to handle the consequences of unitary
evolution of coupled (transmon-like) qubit--resonator quantum systems.

\section*{Acknowledgments}
The authors thank B.M. Terhal and B. Criger for stimulating discussions.
This research is supported by the Early Research Programme of the Netherlands
Organisation for Applied Scientific Research (TNO).
Additional support from the
Top Sector High Tech Systems and Materials is highly appreciated.

\appendix
\section{Definitions and notation}\label{app:A}
\renewcommand{\theequation}{A.\arabic{equation}}
Throughout this paper we adopt units with $\hbar=1$.

For the two data qubit system we use the following definition of the Bell states
\begin{eqnarray}
|\Phi^{\pm}\rangle = \tfrac{1}{2}\sqrt{2}\left(|\mathbf{0, 0}\rangle
\pm |\mathbf{1, 1}\rangle\right), \nonumber \\
|\Psi^{\pm}\rangle = \tfrac{1}{2}\sqrt{2}\left(|\mathbf{0, 1}\rangle
\pm |\mathbf{1, 0}\rangle\right).
\label{eq:Bell}
\end{eqnarray}
The operators $\sigma_z^{[1]}, \sigma_z^{[2]},
\sigma_x^{[1]}$ and $\sigma_x^{[2]}$ are also respectively denoted by
$Z^{[1]}, Z^{[2]}, X^{[1]}, Z^{[2]}$. The superindex refers to the qubit; for the ancillas
the indices $a,b$ will be analogously used.
Below we need the action of the $X$-operators on the basis states 
\begin{eqnarray}
X^{[1]} |\Phi^+\rangle = |\Psi^+\rangle, &\quad& X^{[2]} |\Phi^+\rangle = |\Psi^+\rangle, \nonumber \\
X^{[1]} |\Phi^-\rangle = - |\Psi^-\rangle, &\quad& X^{[2]} |\Phi^-\rangle = |\Psi^-\rangle, \nonumber \\
X^{[1]} |\Psi^+\rangle = |\Phi^+\rangle, &\quad& X^{[2]} |\Psi^+\rangle = |\Phi^+\rangle, \nonumber \\
X^{[1]} |\Psi^-\rangle = - |\Phi^-\rangle, &\quad& X^{[2]} |\Psi^-\rangle = |\Phi^-\rangle.
\label{eq:Xact}
\end{eqnarray}
Because we will consider a $X$ and $Z$ stabilizer circuit, we apply the $Z$-operators
to the Bell states as well and get
\begin{eqnarray}
Z^{[1]} |\Phi^+\rangle = |\Phi^-\rangle, &\quad& Z^{[2]} |\Phi^+\rangle = |\Phi^-\rangle, \nonumber \\
Z^{[1]} |\Phi^-\rangle = |\Phi^+\rangle, &\quad& Z^{[2]} |\Phi^-\rangle = |\Phi^+\rangle, \nonumber \\
Z^{[1]} |\Psi^+\rangle = |\Psi^-\rangle, &\quad& Z^{[2]} |\Psi^+\rangle = -|\Psi^-\rangle, \nonumber \\
Z^{[1]} |\Psi^-\rangle =  |\Psi^+\rangle, &\quad& Z^{[2]} |\Psi^-\rangle = -|\Psi^+\rangle.
\label{eq:Zact}
\end{eqnarray}
A general two-qubit state, usually written as superposition of computational states, can also
 be written as a linear combination of the Bell states, {\it i.e.},
\begin{equation}
|\Psi_{12}\rangle = A_+ |\Phi^+\rangle + A_- |\Phi^-\rangle + B_+ |\Psi^+\rangle + B_- |\Psi^-\rangle,
\label{eq:genstat}
\end{equation}
with
$|A_+|^2+ |A_-|^2 + |B_+|^2 + |B_-|^2 =1 $.

\begin{widetext}
\section{Circuit analyis in terms of Bell states}\label{app:appB}
\renewcommand{\theequation}{B.\arabic{equation}}
\setcounter{equation}{0}
In this appendix we also use the one-qubit states
\begin{equation}
|\bm{\pm}\rangle
= \tfrac{1}{2}\sqrt{2}\left(|\mathbf{0}\rangle
\pm |\mathbf{1}\rangle\right).
\end{equation}
Recall that they are obtained via the Hadamard transformation  on the
computational states
\begin{equation}
\bm{H} \ketz = |\bm{+}\rangle, \qquad
\bm{H} \keto = |\bm{-}\rangle.
\end{equation}
We re-analyze the $X$ and $Z$ stabilizing circuit in terms of Bell states, cf. \cite{Fowler}. 
Explicitly the following steps are done: 
\begin{enumerate}
\item Initialize both ancillas in their ground state. The data qubits are in a general state 
given in (\ref{eq:genstat}). Consequently the initial four-qubit state reads
\begin{eqnarray}
|\psi_1\rangle = \ketz \left(A_+ |\Phi^+\rangle + A_- |\Phi^-\rangle +
B_+ |\Psi^+\rangle + B_-  |\Psi^-\rangle\right) \ketz.
\end{eqnarray}
\item Perform a Hadamard operation on ancilla A 
\begin{equation}
|\psi_2\rangle = |\bm{+}\rangle \left(A_+ |\Phi^+\rangle + A_- |\Phi^-\rangle +
B_+ |\Psi^+\rangle + B_-  |\Psi^-\rangle\right) \ketz.
\end{equation}
\item With ancilla A as control and qubit 1 as target, a CNOT is applied:
\begin{eqnarray}
|\psi_3\rangle &=& \tfrac{1}{2}\sqrt{2} [\ketz  \left(A_+ |\Phi^+\rangle + A_- |\Phi^-\rangle +
B_+ |\Psi^+\rangle + B_-  |\Psi^-\rangle\right) \nonumber \\ &+&
\keto  \left(A_+ |\Psi^+\rangle - A_- |\Psi^-\rangle +
B_+ |\Phi^+\rangle - B_-  |\Phi^-\rangle\right)] \ketz.
\end{eqnarray}
\item A second CNOT is applied, again with ancilla A as control.
Qubit 2, however, is the target:
\begin{eqnarray}
|\psi_4\rangle &=& \tfrac{1}{2}\sqrt{2} [\ketz  \left(A_+ |\Phi^+\rangle + A_- |\Phi^-\rangle +
B_+ |\Psi^+\rangle + B_-  |\Psi^-\rangle\right) \nonumber \\ &+&
\keto  \left(A_+ |\Phi^+\rangle - A_- |\Phi^-\rangle +
B_+ |\Psi^+\rangle - B_-  |\Psi^-\rangle\right)] \ketz \\
&=& |\bm{+}\rangle  \left(A_+ |\Phi^+\rangle + 
B_+ |\Psi^+\rangle \right) \ketz +
|\bm{-}\rangle  \left(A_- |\Phi^-\rangle +
B_-|\Psi^-\rangle\right)] \ketz. \nonumber
\end{eqnarray}
\item A CNOT is applied with data qubit 1 as control and ancilla B as target.
We omit the intermediate result $|\psi_5\rangle$  because it is more convenient
to combine this operation with the next step.
\item A CNOT is applied with data qubit 2 as control and ancilla B  as target.
The two CNOTs yield for the relevant states
\begin{equation}
|\Phi^{\pm}\rangle \ketz \rightarrow |\Phi^{\pm}\rangle \ketz,  \quad
|\Psi^{\pm}\rangle \ketz \rightarrow |\Psi^{\pm}\rangle \keto. 
\end{equation}
Therefore, the resulting four-qubit state reads
\begin{equation}
|\psi_6\rangle =
A_+ |\bm{+}\rangle  |\Phi^+\rangle \ketz + B_+ |\bm{+}\rangle |\Psi^+\rangle \keto 
+ A_- |\bm{-}\rangle  |\Phi^-\rangle \ketz + B_- |\bm{-}\rangle |\Psi^-\rangle \keto.
\end{equation}
\item A Hadamard is performed on ancilla A; we obtain
\begin{equation}
|\psi_7\rangle =
A_+ \ketz |\Phi^+\rangle \ketz + B_+ \ketz  |\Psi^+\rangle \keto 
+ A_- \keto |\Phi^-\rangle \ketz + B_- \keto |\Psi^-\rangle \keto.
\label{eq:Had4}
\end{equation}
\item Both ancillas are eventually measured in their standard basis. The concomitant measurement
operators are given by
\begin{eqnarray}
P_{11} = \ketz \braz \otimes \mathcal{I}_{12} \otimes \ketz \braz, &\quad&
P_{-11} = \keto \brao \otimes \mathcal{I}_{12} \otimes \ketz \braz, \nonumber \\
P_{1-1} = \ketz \braz \otimes \mathcal{I}_{12} \otimes \keto \brao, &\quad&
P_{-1-1} = \keto \brao \otimes \mathcal{I}_{12} \otimes \keto \brao,
\end{eqnarray}
with  probabilities for the outcomes $\pm 1, \pm 1$
\begin{eqnarray}
p(1,1) =  |A_+|^2, &\quad& p(1,-1)= |B_+|^2, \nonumber \\
p(-1,1) =  |A_-|^2, &\quad& p(-1,-1) = |B_-|^2.
\end{eqnarray}
The corresponding final two-qubit states are the Bell states
\begin{eqnarray}
\text{syndrome} \,(1,1):   |\Phi^+\rangle, &\qquad&
\text{syndrome}\,(1,-1):   |\Psi^+\rangle, \nonumber \\
\text{syndrome} \,(-1,1):  |\Phi^-\rangle, &\qquad&
\text{syndrome} \,(-1,-1):  | \Psi^-\rangle.
\end{eqnarray}
\end{enumerate}

\section{Basis states and matrix elements}\label{app:appC}
\renewcommand{\theequation}{C.\arabic{equation}}
\setcounter{equation}{0}
\subsection{First excitation level}
In this subspace, we choose the basis states
\begin{eqnarray}
|e_1\rangle = |1,0,0,0\rangle \otimes |\bm{0},\bm{0}\rangle  \otimes |\bm{0}, \bm{0}\rangle,&& \quad
|e_2\rangle = |0,1,0,0\rangle \otimes |\bm{0},\bm{0}\rangle  \otimes |\bm{0}, \bm{0}\rangle, \nonumber \\
|e_3\rangle = |0,0,1,0\rangle \otimes |\bm{0},\bm{0}\rangle  \otimes |\bm{0}, \bm{0}\rangle,&& \quad
|e_4\rangle = |0,0,0,1\rangle \otimes |\bm{0},\bm{0}\rangle  \otimes |\bm{0}, \bm{0}\rangle, \nonumber \\
|e_5\rangle = |0,0,0,0\rangle \otimes |\bm{1},\bm{0}\rangle  \otimes |\bm{0}, \bm{0}\rangle,&& \quad
|e_6\rangle = |0,0,0,0\rangle \otimes |\bm{0},\bm{1}\rangle  \otimes |\bm{0}, \bm{0}\rangle, \nonumber \\
|e_7\rangle = |0,0,0,0\rangle \otimes |\bm{0},\bm{0}\rangle  \otimes |\bm{1}, \bm{0}\rangle,&& \quad
|e_8\rangle = |0,0,0,0\rangle \otimes |\bm{0},\bm{0}\rangle  \otimes |\bm{0}, \bm{1}\rangle.
\end{eqnarray}
The matrix elements of the Hamiltonian in this subspace can be readily calculated from
\begin{equation}
\mathcal{H}_{ij} = \langle e_i | H | e_j \rangle, \qquad i,j=1,2, \dots 8.
\end{equation}
For $i \le j$ the nonzero elements are
\begin{eqnarray}
\mathcal H_{11} &=& E_0 + \omega_1, \quad \mathcal H_{15} = g_{11}, \quad \mathcal H_{17} = g_{1a}, \nonumber \\
\mathcal H_{22} &=& E_0 + \omega_2, \quad \mathcal H_{26} = g_{22}, \quad \mathcal H_{27} = g_{2a}, \nonumber \\
\mathcal H_{33} &=& E_0 + \omega_3, \quad \mathcal H_{35} = g_{31}, \quad \mathcal H_{38} = g_{3b}, \nonumber \\
\mathcal H_{44} &=& E_0 + \omega_4, \quad \mathcal H_{46} = g_{42}, \quad \mathcal H_{48} = g_{4b}, \\
\mathcal H_{55} &=& E_0 + \omega'_1, \quad \mathcal H_{66} = E_0 + \omega'_2,  \quad
\mathcal H_{77} = E_0 + \omega_a, \quad \mathcal H_{88} = E_0 + \omega_b, \nonumber
\end{eqnarray}
whereas the other ones follow by symmetry $\mathcal H_{ij}= \mathcal H_{ji}$.

\subsection{Second excitation level}
We proceed to the second excitation level which has dimension thirty-two. Its basis is chosen as
\begin{eqnarray}
|f_1\rangle = |2,0,0,0\rangle \otimes |\bm{0},\bm{0}\rangle  \otimes |\bm{0}, \bm{0}\rangle,&& \quad
|f_2\rangle = |1,1,0,0\rangle \otimes |\bm{0},\bm{0}\rangle  \otimes |\bm{0}, \bm{0}\rangle, \nonumber \\
|f_3\rangle = |1,0,1,0\rangle \otimes |\bm{0},\bm{0}\rangle  \otimes |\bm{0}, \bm{0}\rangle,&& \quad
|f_4\rangle = |1,0,0,1\rangle \otimes |\bm{0},\bm{0}\rangle  \otimes |\bm{0}, \bm{0}\rangle, \nonumber \\
|f_5\rangle = |0,2,0,0\rangle \otimes |\bm{0},\bm{0}\rangle  \otimes |\bm{0}, \bm{0}\rangle,&& \quad
|f_6\rangle = |0,1,1,0\rangle \otimes |\bm{0},\bm{0}\rangle  \otimes |\bm{0}, \bm{0}\rangle, \nonumber \\
|f_7\rangle = |0,1,0,1\rangle \otimes |\bm{0},\bm{0}\rangle  \otimes |\bm{0}, \bm{0}\rangle,&& \quad
|f_8\rangle = |0,0,2,0\rangle \otimes |\bm{0},\bm{0}\rangle  \otimes |\bm{0}, \bm{0}\rangle \nonumber \\
|f_9\rangle = |0,0,1,1\rangle \otimes |\bm{0},\bm{0}\rangle  \otimes |\bm{0}, \bm{0}\rangle,&& \quad
|f_{10}\rangle = |0,0,0,2\rangle \otimes |\bm{0},\bm{0}\rangle  \otimes |\bm{0}, \bm{0}\rangle \nonumber \\
|f_{11}\rangle = |1,0,0,0\rangle \otimes |\bm{1},\bm{0}\rangle  \otimes |\bm{0}, \bm{0}\rangle,&& \quad
|f_{12}\rangle = |1,0,0,0\rangle \otimes |\bm{0},\bm{1}\rangle  \otimes |\bm{0}, \bm{0}\rangle \nonumber \\
|f_{13}\rangle = |1,0,0,0\rangle \otimes |\bm{0},\bm{0}\rangle  \otimes |\bm{1}, \bm{0}\rangle,&& \quad
|f_{14}\rangle = |1,0,0,0\rangle \otimes |\bm{0},\bm{0}\rangle  \otimes |\bm{0}, \bm{1}\rangle \nonumber \\
|f_{15}\rangle = |0,1,0,0\rangle \otimes |\bm{1},\bm{0}\rangle  \otimes |\bm{0}, \bm{0}\rangle,&& \quad
|f_{16}\rangle = |0,1,0,0\rangle \otimes |\bm{0},\bm{1}\rangle  \otimes |\bm{0}, \bm{0}\rangle  \\
|f_{17}\rangle = |0,1,0,0\rangle \otimes |\bm{0},\bm{0}\rangle  \otimes |\bm{1}, \bm{0}\rangle,&& \quad
|f_{18}\rangle = |0,1,0,0\rangle \otimes |\bm{0},\bm{0}\rangle  \otimes |\bm{0}, \bm{1}\rangle \nonumber \\
|f_{19}\rangle = |0,0,1,0\rangle \otimes |\bm{1},\bm{0}\rangle  \otimes |\bm{0}, \bm{0}\rangle,&& \quad
|f_{20}\rangle = |0,0,1,0\rangle \otimes |\bm{0},\bm{1}\rangle  \otimes |\bm{0}, \bm{0}\rangle \nonumber \\
|f_{21}\rangle = |0,0,1,0\rangle \otimes |\bm{0},\bm{0}\rangle  \otimes |\bm{1}, \bm{0}\rangle,&& \quad
|f_{22}\rangle = |0,0,1,0\rangle \otimes |\bm{0},\bm{0}\rangle  \otimes |\bm{0}, \bm{1}\rangle \nonumber \\
|f_{23}\rangle = |0,0,0,1\rangle \otimes |\bm{1},\bm{0}\rangle  \otimes |\bm{0}, \bm{0}\rangle,&& \quad
|f_{24}\rangle = |0,0,0,1\rangle \otimes |\bm{0},\bm{1}\rangle  \otimes |\bm{0}, \bm{0}\rangle \nonumber \\
|f_{25}\rangle = |0,0,0,1\rangle \otimes |\bm{0},\bm{0}\rangle  \otimes |\bm{1}, \bm{0}\rangle,&& \quad
|f_{26}\rangle = |0,0,0,1\rangle \otimes |\bm{0},\bm{0}\rangle  \otimes |\bm{0}, \bm{1}\rangle \nonumber \\
|f_{27}\rangle = |0,0,0,0\rangle \otimes |\bm{1},\bm{1}\rangle  \otimes |\bm{0}, \bm{0}\rangle,&& \quad
|f_{28}\rangle = |0,0,0,0\rangle \otimes |\bm{1},\bm{0}\rangle  \otimes |\bm{1}, \bm{0}\rangle \nonumber \\
|f_{29}\rangle = |0,0,0,0\rangle \otimes |\bm{1},\bm{0}\rangle  \otimes |\bm{0}, \bm{1}\rangle,&& \quad
|f_{30}\rangle = |0,0,0,0\rangle \otimes |\bm{0},\bm{1}\rangle  \otimes |\bm{1}, \bm{0}\rangle \nonumber \\
|f_{31}\rangle = |0,0,0,0\rangle \otimes |\bm{0},\bm{1}\rangle  \otimes |\bm{0}, \bm{1}\rangle,&& \quad
|f_{32}\rangle = |0,0,0,0\rangle \otimes |\bm{0},\bm{0}\rangle  \otimes |\bm{1}, \bm{1}\rangle. \nonumber
\end{eqnarray}
The matrix elements in the second excitation subspace follow analogously as
\begin{equation}
\mathsf{H}_{kl} = \langle f_k | H | f_l \rangle, \qquad k,l=1,2, \dots, 32.
\end{equation}
For the diagonal ones we get
\begin{eqnarray}
\mathsf H_{11} &=& E_0 +2\omega_1, \quad \mathsf H_{22}=E_0+\omega_1+\omega_2, \quad
\mathsf H_{33} = E_0+\omega_1+\omega_3, \quad \mathsf H_{44} = E_0 +\omega_1+\omega_4, \nonumber \\
\mathsf H_{55} &=& E_0 +2\omega_2, \quad \mathsf H_{66}=E_0+\omega_2+\omega_3, \quad
\mathsf H_{77} = E_0+\omega_2+\omega_4, \quad \mathsf H_{88} = E_0 + 2 \omega_3, \nonumber \\
\mathsf H_{99} &=& E_0 +\omega_3 + \omega_4, \quad \mathsf H_{10-10}=E_0+ 2 \omega_4, \quad
\mathsf H_{11-11} = E_0+\omega_1+\omega'_1, \nonumber \\
\mathsf H_{12-12} &=& E_0 +\omega_1+\omega'_2, \quad \mathsf H_{13-13}=E_0+\omega_1+\omega_a, \quad
\mathsf H_{14-14} = E_0+\omega_1+\omega_b, \nonumber \\
\mathsf H_{15-15} &=& E_0 +\omega_2+\omega'_1, \quad \mathsf H_{16-16}=E_0+\omega_2+\omega'_2, \quad
\mathsf H_{17-17} = E_0+\omega_2+\omega_a, \nonumber \\
\mathsf H_{18-18} &=& E_0 +\omega_2+\omega_b, \quad \mathsf H_{19-19}=E_0+\omega_3+\omega'_1, \quad
\mathsf H_{20-20} = E_0+\omega_3+\omega'_2, \nonumber \\
\mathsf H_{21-21} &=& E_0 +\omega_3+\omega_a, \quad \mathsf H_{22-22}=E_0+\omega_3+\omega_b, \quad
\mathsf H_{23-23} = E_0+\omega_4+\omega'_1, \nonumber \\
\mathsf H_{24-24} &=& E_0 +\omega_4+\omega'_2, \quad \mathsf H_{25-25}=E_0+\omega_4+\omega_a, \quad
\mathsf H_{26-26} = E_0+\omega_4+\omega_b, \nonumber \\
\mathsf H_{27-27} &=& E_0 +\omega'_1+\omega'_2, \quad \mathsf H_{28-28}=E_0+\omega'_1+\omega_a, \quad
\mathsf H_{29-29} = E_0+\omega'_1+\omega_b, \nonumber \\
\mathsf H_{30-30} &=& E_0 +\omega'_2+\omega_a, \quad \mathsf H_{31-31}=E_0+\omega'_2+\omega_b, \quad
\mathsf H_{32-32} = E_0+\omega_a+\omega_b. 
\end{eqnarray}
The nonzero matrix  elements $\mathsf H_{kl}$ with $k<l$ are given by
\begin{eqnarray}
\mathsf H_{1-11} &=& \sqrt{2} g_{11}, \quad \mathsf H_{1-13}= \sqrt{2} g_{1a}, \nonumber \\
\mathsf H_{2-12} &=&  g_{22}, \quad \mathsf H_{2-13}=  g_{2a}, \quad 
\mathsf H_{2-15} = g_{11}, \quad \mathsf H_{2-17}= g_{1a}, \nonumber \\ 
\mathsf H_{3-11} &=& g_{31}, \quad \mathsf H_{3-14}= g_{3b}, \quad 
\mathsf H_{3-19} =  g_{11}, \quad \mathsf H_{3-21}=  g_{1a}, \nonumber \\
\mathsf H_{4-12} &=& g_{42}, \quad \mathsf H_{4-14}= g_{4b}, \quad 
\mathsf H_{4-23} =  g_{11}, \quad \mathsf H_{4-25}=  g_{1a}, \nonumber \\
\mathsf H_{5-16} &=& \sqrt{2} g_{22}, \quad \mathsf H_{5-17}= \sqrt{2} g_{2a}, \nonumber \\
\mathsf H_{6-15} &=& g_{31}, \quad \mathsf H_{6-18}= g_{3b}, \quad 
\mathsf H_{6-20} =  g_{22}, \quad \mathsf H_{6-21}=  g_{2a}, \nonumber \\
\mathsf H_{7-16} &=& g_{42}, \quad \mathsf H_{7-18}= g_{4b}, \quad 
\mathsf H_{7-24} =  g_{22}, \quad \mathsf H_{7-25}=  g_{2a}, \nonumber \\
\mathsf H_{8-19} &=& \sqrt{2} g_{31}, \quad \mathsf H_{8-22}= \sqrt{2} g_{3b}, \nonumber \\
\mathsf H_{9-20} &=& g_{42}, \quad \mathsf H_{9-22}= g_{4b}, \quad 
\mathsf H_{9-23} =  g_{31}, \quad \mathsf H_{9-26}=  g_{3b}, \nonumber \\
\mathsf H_{10-24} &=& \sqrt{2} g_{42}, \quad \mathsf H_{10-26}= \sqrt{2}g_{4b}, \quad 
\mathsf H_{11-28} =  g_{1a}, \nonumber \\
\mathsf H_{12-27} &=&  g_{11}, \quad \mathsf H_{12-30}= g_{1a}, \quad 
\mathsf H_{13-28} =  g_{11}, \nonumber \\
\mathsf H_{14-29} &=& g_{11}, \quad \mathsf H_{14-32}= g_{1a}, \quad 
\mathsf H_{15-27} =  g_{22}, \quad \mathsf H_{15-28}=  g_{2a}, \nonumber \\
\mathsf H_{16-30} &=& g_{2a}, \quad \mathsf H_{17-30}= g_{22}, \quad 
\mathsf H_{18-31} =  g_{22}, \quad \mathsf H_{18-32}=  g_{2a}, \nonumber \\
\mathsf H_{19-29} &=&  g_{3b}, \quad \mathsf H_{20-27}= g_{31}, \quad 
\mathsf H_{20-31} =  g_{3b}, \nonumber \\
\mathsf H_{21-28} &=&  g_{31}, \quad \mathsf H_{21-32}= g_{3b}, \quad 
\mathsf H_{22-29} =  g_{31}, \nonumber \\
\mathsf H_{23-27} &=&  g_{42}, \quad \mathsf H_{23-29}= g_{4b}, \quad 
\mathsf H_{24-31} =  g_{4b}, \nonumber \\
\mathsf H_{25-30} &=&  g_{42}, \quad \mathsf H_{25-32}= g_{4b}, \quad 
\mathsf H_{26-31} =  g_{42} 
\end{eqnarray}
and those with $k > l$ follow by symmetry $\mathsf H_{kl} =\mathsf H_{lk} $.
\end{widetext}

\end{document}